\documentclass[preprints,article,accept,oneauthor,pdftex]{mdpi}
\firstpage{1} 
\pubvolume{xx}
\issuenum{1}
\articlenumber{xx}
\pubyear{2019}
\copyrightyear{2019}
\history{\large Paper submitted to the review "ENTROPY" on 31 October, 2019.}

\Title{Application of the second law to the atmosphere: impacts of the third-law definition for the moist-air entropy.}

\Author{Pascal Marquet $^{1}$\orcidA{}}

\AuthorNames{Pascal Marquet}

\address{$^{1}$ \quad M\'et\'eo-France, CNRM-CNRS UMR-3589; GMAP, 42 avenue G. Coriolis, 31057 Toulouse CEDEX 01. France [1]; pascal.marquet@meteo.fr}

\corres{Correspondence: pascal.marquet@meteo.fr; Tel.: +33-5-61-07-84-38}

\abstract{Calculations of entropy fluxes and production rate have been evaluated with some success to study atmospheric processes.
However, recurring questions arise as to how best to take into account entropy flux due to radiation, for example.
This article raises another kind of question: how to define the entropy of the atmosphere itself, which is composed of variable proportions of dry air (nitrogen, oxygen, argon, etc.) and water (vapour, liquid, ice).
The specific values of the entropy for such a variable composition system depend on the reference values of its components.
Most of the current definitions are based on entropies set at zero for dry air and liquid water at zero degrees Celsius.
Differently, the third law of thermodynamics assumes that the entropy of all species cancels out for the more stable solid state at the zero of absolute temperatures.
In this paper, we analyze the possible consequences of this absolute definition of entropy of moist air on the calculation of entropy fluxes.
The impacts of moisture are significant and these new calculation methods seem to be able to modify the budgets of atmospheric entropy, with possible impacts on the nature of the equilibrium of the atmosphere resulting from entropic imbalances induced by radiations.
}


\usepackage{hyperref}

\begin{document}


 \section{Introduction} 



The pioneering works of Onsager, Eckart, von Meixner, Prigogine, van Mieghem, de~Groot, von Hasse, Denbigh, Mazur and Glansdorff
\cite{Onsager_1931_II} 
\cite{Eckart_1940_I} \cite{Eckart_1940_II} 
\cite{Meixner_1941} \cite{Meixner_1942} \cite{Meixner_1943}
\cite{Prigogine_1947} \cite{Mieghem_1947} \cite{deGroot_1951}  \cite{Hasse_II_1951}  \cite{Denbigh_1952} \cite{Mazur_1952} \cite{Glansdorff_1953} are now summarized in the books \cite{Glansdorff_Prigogine_1971} \cite{deGroot_Mazur_86}.
Their common goal was to study the equation expressing the local equilibrium for the entropy of the atmosphere, considered as moist air made of a mixture of perfect gases (dry air, water vapour, liquid water droplets and ice crystals).

To do this, it is necessary to make the hypothesis of local thermodynamic equilibrium, i.e. that air in motion and out of equilibrium remains sufficiently close to equilibrium so that the state functions (such as energy and entropy) can be expressed with local average variables such as temperature, pressure and concentrations of the different components of moist air.

The local entropy equation are written in the following general form
\begin{align}
 \rho \, \frac{d s }{d t}
 =
 \frac{\partial \, \rho \, s }{\partial t}
 +
 \vec{\nabla} . 
 \left( \rho \, s \, \vec{u} \right)
 & = \:
 \rho \, \frac{d_e s }{d t}
 \: + \:
 \rho \, \frac{d_i s }{d t} \: ,
 \label{Eq_Entropy_1} \\
 \rho \, \frac{d_e s }{d t}
 & = \:
 - \, \vec{\nabla} . \vec{J}_R
 - \, \vec{\nabla} . \vec{J}_s \: ,
 \label{Eq_Entropy_1e} \\
 \rho \, \frac{d_i s }{d t}
 & = \:
 \sigma \: .
 \label{Eq_Entropy_1i}
\end{align}
The rate of "{\em external\/}" entropy change $\rho \,{d_e s }/{d t}$ can be positive or negative.
It is generated by the entropy flux due to short-wave and long-wave radiations ($\vec{J}_R$) and the entropy flux of matter ($\vec{J}_s$).
The rate of "{\em internal\/}" entropy change  $\rho  \,{d_i s }/{d t} = \sigma \ge 0$ is called the "{\em rate of entropy production\/}".
It is expected to be positive due to the second law of thermodynamics.

It is possible to simply illustrate this separation into a flux $\vec{J}_s$ and a positive production term $\sigma$ for the particular case of thermal conduction with a flux of energy $\vec{F}_T$.
The second law of thermodynamics implies that
\begin{align}
 \rho \, \left. \frac{ds}{dt} \right|_{cond.}
 \: = \: \frac{\left. \dot{q} \right|_{T}}{T}
 \: = \: 
 \: \frac{ - \, \vec{\nabla} . \, \vec{F}_T }{T}
 \: = \:
 \: - \, \vec{\nabla} . \left( \frac{ \vec{F}_T }{T} \right)
 \: + \: \vec{F}_T \, . \vec{\nabla} \left( \frac{1}{T}\right)
 \label{Eq_Entropy_1_cond} \: ,
\end{align}
where the last two terms are another way of writing 
$- \, \vec{\nabla} . \, \vec{F}_T / {T}$.
The term $\vec{F}_T / T$ is considered a contribution to $\vec{J}_s$, while $\vec{F}_T \, . \vec{\nabla} ({1}/{T})$ is a contribution to $\sigma$ since it is positive for the usual case of energy fluxes that operate in counter-gradients (i.e. via a Fick's law with 
$\vec{F}_T = - a \: \vec{\nabla} T$ with $a>0$).

The interest of making such a separation in terms of entropy flow and entropy production is to be able to analyze whether the earth system is in maximum entropy production mode (MEP), or not.

Differently, Mobbs \cite{Mobbs_1986} recalled that Paltridge \citep{Paltridge_1975} "{\em hypothesized that the climate system could be constrained to operate at near a local minimum of the entropy exchange rate and used this closure condition with remarkable success in an energy balanced atmosphere-ocean climate model\/}."
In fact, these two approaches are similar since Paltridge's approach is to focus on the quantity 
\begin{align}
 \rho \, \frac{d_e s }{d t}
 & = \:
  - \, \vec{\nabla} . \left( \vec{J}_s + \vec{J}_R \right)
 \: = \:
 \rho \, \frac{d s }{d t}
 \: - \: \sigma \: ,
 \label{Eq_Entropy_2}
\end{align}
and then consider that $ds/dt$ is zero on long-term and global average, to deduce that $d_e s/dt$ is minimum if $\sigma$ is maximum.
The advantage of this method is that it is easier to calculate the integral on the atmosphere of this divergence than the term $\sigma$, this integral being equal to the integral of the vertical components of $\vec{J}_s$ and $\vec{J}_R$ at the surface and the top of the atmosphere, whereas it would be necessary to know $\sigma$ at each point in the atmosphere.
 
While Paltridge's results were promising, many questions remained unresolved and even created controversy.
In particular, the question of calculating entropy flux ($\vec{J}_R$) due to solar and infrared radiation still seems to be a matter of debate.
This calculation was first done by simply adding the upward fluxes $SW/T_{sun}$ and $LW/T_{earth}$ for an energy balance of $LW \approx -\, SW \approx 239$~W~m${}^{-2}$, resulting in an imbalance and destruction of the entropy at the rate of about $239 \: (1/5778 \, - \, 1/255) \approx -0.9$~W~m${}^{-2}$~K${}^{-1}$.
These kinds of simple calculations were made in 
\citeauthor{Peixoto_al_1991} \cite{Peixoto_al_1991} and others (\cite{Pauluis_Held_2002_I},
\cite{Ozawa_2003}, 
\cite{Kleidon_Lorenz_2005}, 
\cite{Fraedrich_Lunkeit_2008},
\cite{Kleidon_2010},
\cite{Pascale_al_2011},
\cite{Pascale_al_ClimDyn_2012},
\citep{Bannon_Lee_2017}).

Differently, \citeauthor{Essex_1984c} \citep{Essex_1984c} applied the formula of \citeauthor{Planck_1914} \cite{Planck_1914} which indicates that the flux of entropy emitted by a black body at the temperature $T$ is $(4/3) \, \sigma_B \: T^3 = (4/3)E/T$, where $E$ is the energy emitted by the black body and $\sigma_B$ the Stefan-Boltzmann constant.
These calculations based on Planck's formula have been used in \cite{Mobbs_1986} and then extended by \citeauthor{Stephens_OBrien_1993} \cite{Stephens_OBrien_1993} and \citeauthor{Goody_Abdou_1996} \cite{Goody_Abdou_1996} to the case of the radiation reflected by the earth's surface, with in this case an entropy flux that is equal to $(4/3) \: \sigma_B \: T^3_{sun} \: \chi(u)$ where $u = \alpha \: \cos(\beta) \: (\Omega_0/\pi)$ depends on the  
albedo of earth $\alpha$, the solid angle subtended by the sun $\Omega_0$ and the zenith angle of the sun $\beta$.
Several papers have subsequently used this Planck's coefficient $(4/3)$ for radiation entropy (\cite{Pelkowski_2012}, \cite{Bannon_2015}) and the function $\chi(u)$ (\cite{Wu_Liu_2010}).
The rejection of the $4/3$ factor in the publications cited above is severely judged in (\cite{Pelkowski_2012}): "{\em the reasons adduced being either ad hoc or mere hocus-pocus.\/}"

Both the $4/3$ coefficient (which applies to solar and infrared fluxes emitted by the sun and earth) and the function $\chi(u) \approx u \, [\: 0.9652 - 0.278 \: \ln(u) \: ]$ for $u <0.1$ (which applies to reflected solar fluxes) induce a rate of destruction of entropy at the TOA of about $-1.3$~W~m${}^{-2}$~K${}^{-1}$, which is $30$~\% greater than previous evaluations \cite {Wu_Liu_2010}.

To overcome the difficulties related to uncertainties in entropy fluxes due to radiation, as well as the difficulty of directly calculating the production term $\sigma$, it must be possible to directly calculate the entropy $s$ from the atmosphere, and thus study Equation~(\ref{Eq_Entropy_1}) for $ds/dt$ differently.
More precisely, the objective would be to calculate the two terms 
${\partial ( \rho \, s ) }/{\partial t}$
and
$ \vec{\nabla} . \left( \rho \, s \, \vec{u} \right)$
in the left-hand side of Equation~(\ref{Eq_Entropy_1}).

This proposal corresponds to the remarks made by \citeauthor{Golitsyn_Mokhov_1978} \cite{Golitsyn_Mokhov_1978}:
``It would be of great interest to compute the quantities $\partial s/\partial t$, $ds_e/dt$ and $ds_i/dt$ as global characteristics in large three-dimensional numerical models.
Verification of the behaviour of these quantities with time during the calculations may serve as not only an additional criterion of the correctness of operation of the scheme as well as a means of checking the conservation of mass, total kinetic energy, or momentum, but it also could clarify a great deal in the physics of atmospheric processes''.
These calculations must be possible because the specific entropy ($s$) is a state function that can be calculated from the basic thermodynamic variables, such as temperature, pressure and concentrations of different phases of water.
However, these calculations have been performed by generating another source of uncertainty, which is related to the different choices made to measure the entropy of moist air.

Some papers have considered only the variation of entropy with temperature, with possible effects related to phase changes in water either neglected or being taken into account through diabatic sources of temperature.
(\cite{Nicolis_Nicolis_1980}, 
\cite{Ozawa_2003},
\cite{Pascale_al_2011}). 
Several papers have used the definition of entropy for dry air, including the impacts of both temperature and pressure but without direct impact from water vapour, liquid water or ice
(\cite{Johnson_1997},
\cite{Woollings_Thuburn_2006},
\cite{Akmaev_2008},
\cite{Liu_al_2011},
\cite{Gassmann_al_2015}).
Some other papers have defined the entropy of moist air without applying the recommendations related to the third law of thermodynamics, formerly the  Nernst's heat theorem
(\citep{Emanuel94},
\cite{Noda_Tokioka_1983},
\cite{Goody_2000},
\cite{Pauluis_Held_2002_I},
\cite{Liu_Liu_2008},
\cite{Pauluis_al_2010},
\cite{Feistel_al_2010},
\cite{Pauluis_2011},
\cite{Mrowiec_al_2016},
\cite{Pauluis_2016}
\cite{Feistel_2018},
\cite{Feistel_2019}).
In these papers it is assumed that the entropy of dry air and liquid water is equal to zero at the temperature of zero Celcius.
Differently, the third law requires that the entropy of the most stable solid form of all chemical components must be set to zero at the absolute zero of the temperature ($0$~K or $-273.15$~K). 

A surprising fact is that the two sources of uncertainty are related, as Planck clearly explained when he wrote his two treatises on radiation in 1913 \cite{Planck_1914} and on thermodynamics in 1917 \cite{Planck_1917}.
This is due to the fact that Planck has devoted his life to the applications of thermodynamics and the second law, in particular for the study of the properties of radiation \cite{Muller_2008}.

Indeed, \citeauthor{Planck_1914} \cite{Planck_1914} (pages 141-142)
define the entropy of radiation starting from the Boltzmann formula 
$S = k_B \: \ln(W) + const.$ (\cite{Boltzmann_1877}, \cite{Planck_1901}), which is "{\em an equation which determines the general way in which the entropy depends on the probability.\/}"
And this "{\em universal constant of integration $k_B$ is the same for a terrestrial as for a cosmic system, and its value, having been determined for the former, will remain valid for the latter.\/}"
Moreover, the "{\em second additive constant of integration (const.) may, without any restriction as regards generality, be included as a constant multiplier in the quantity $W$, (..) so that the equation reduces to $S = k_B \: \ln(W)$.\/}"
Then \citeauthor{Planck_1914} "{\em assigns a definite absolute value to the entropy $S$. 
This is a step of fundamental importance, which can be justified only by its consequences. 
(...) this step leads necessarily to the "hypothesis of quanta" and moreover it also leads, as regards radiant heat, to a definite law of distribution of energy of black radiation, and, as regards heat energy of bodies, to Nernst's heat theorem.\/}"
\citeauthor{Planck_1914}  computes (pages 72-74, 76) the entropy of black-body radiation in a volume $V$ by 
$dS(T,V) = (dU + p\,dV)/T$, 
with $u = U/V = \sigma_B \: T^4$ and $p = u/3$ leading to
$dS = d[\: (4/3) \: \sigma_B \: T^3 \: V \: ]$ and, by integration, to
$S = (4/3) \: \sigma_B \: T^3 \: V$
up to an additive constant.
He finally explains that in "{\em this equation the additive constant is determined by a choice that readily suggests itself, so that at the zero of the absolute scale of temperature, that is to say, when $u$ vanishes, $S$ shall become zero. 
From this the entropy of unit volume or the volume density of the entropy of black radiation is obtained, $s = S/V = 4/3 \: \sigma_B \: T^3$.\/}"

All these remarks by \citeauthor{Planck_1914} indicate that, for the study of the budget of  entropy for the atmosphere, it is necessary both to use the $(4/3) E/T$ law for the entropy flux of radiation and to calculate the entropy of the atmosphere by setting to zero those of the most stable crystals of nitrogen, oxygen, argon and water ice at the temperature of zero Kelvin.    

Beyond a simple problem of choice that Planck describes as not arbitrary, taking this or that value for the dry air and water vapour entropies has physical consequences that \citeauthor{Richardson_1922} \cite{Richardson_1922} described in these terms (page~159-160, about the impact of evaporation processes on the fluxes of entropy): 
"{\em what energy and entropy are to be ascribed to unit mass of the incoming substance? 
As there is an arbitrary constant of integration in the entropy, we must ask what would be the effect of an increase in this constant for the incoming water.
Approximations are not here permissible, for the constant might be made indefinite large.\/}"



Therefore, the primary objective of this paper is to compute accurately the specific entropy of moist air ($s$) by using the third law of thermodynamics, and to evaluate its surface fluxes via the integral of the divergence term
$ - \, \vec{\nabla} . \left( \rho \, s \, \vec{u} \right)$ 
in Equation~(\ref{Eq_Entropy_1}), written in the form
\begin{align}
 - \: \vec{\nabla} . 
      \left( \rho \, s \, \vec{u} \right)
 - \, \vec{\nabla} . \vec{J}_R
 & = \:
  \frac{\partial \, \rho \, s }{\partial t}
 + \, \vec{\nabla} . \vec{J}_s
 - \, \sigma 
 \label{Eq_Entropy_3} \: .
\end{align}

The final goal would be to be able to assess how much the total atmospheric entropy balance has changed if this new surface entropy flux is taken into account, with $s$ computed with the third-law values and $\vec{J}_R$ evaluated with the factor $(4/3)$ for the entropy flux of radiation.
The last two terms $- \, \vec{\nabla} . \vec{J}_s$ ans $\sigma$ will not be evaluated in this study, as in Paltridge's approach. 

The rest of the paper is organized as follows.
The third-law value of the moist-air entropy is computed in section~ \ref{Entropy_atmosphere} and in Appendix~A.
Section~\ref{Entropy_surf_fluxes} presents the new expressions for the turbulent surface fluxes of the third-law entropy of the atmosphere.
The new global entropy budget of the atmosphere is calculated in the section~\ref{Results_ARPEGE} for seasonal averages of the French ARPEGE NPW model.
More local results are shown in section~\ref{Results_LES_HIGHTune} from the outputs of three simulations with the Meso-NH model in LES mode, and then in section~\ref{Results_EBEX} for data observed during the EBEX-2000 campaign.
The major findings and future research are summarized in section~\ref{Conclusion}.

\section{The entropy of the atmosphere} 
\label{Entropy_atmosphere}

The dry-air ($s_{dry}$) and absolute moist-air ($s_{abs}$) entropies are defined by Equations~(\ref{eq_s_theta_dry}) and (\ref{eq_s_thetas}), leading to 
\begin{align}
s_{dry} \: - \: s_{d}^0 & = \: c_{pd} \: 
    \ln\left(\frac{\theta}{T_0}\right) 
\label{eq_s_theta_dry_plot} \: , \\
 s_{abs} \: - \: s_{d}^0 & \: = \: 
    c_{pd} \: 
     \ln\left(\frac{\theta_s}{T_0}\right) \: , 
  \label{eq_s_thetas_plot}
\end{align}
where $\theta$ and $\theta_s$ are given by Equations~(\ref{eq_theta}) and (\ref{eq_thetas}).
See Appendix A for explanations of the different terms and their numerical values, in particular for the reference entropy of dry air $s_{d}^0$.

The first- and second-order approximations for the absolute moist-air entropy are defined by
$s_{abs/1} = c_{pd} \ln\left[\, {({\theta}_{s})}_1 / T_0\, \right] 
+ s_{d}^0$
and 
$s_{abs/2} = c_{pd} \ln\left[\, {({\theta}_{s})}_2 / T_0\, \right] 
+ s_{d}^0$,
with ${({\theta}_{s})}_1$ and ${({\theta}_{s})}_2$ given by Equations ~(\ref{eq_thetas1}) to (\ref{eq_thetas2_bis}), leading to
\begin{align}
 s_{abs/1} - s_{d}^0
   & = \: c_{pd} \, 
   \ln\left(\frac{\theta}{T_0}\right)
    - \: \frac{L_v}{T} \: q_l
 \: - \: \frac{L_s}{T} \: q_i
 \: + \: c_{pd} \: \Lambda \: q_t
\label{eq_s1_plot} \: , \\
 s_{abs/2}  -  s_{d}^0
   & = \: c_{pd} \, 
   \ln\left(\frac{\theta}{T_0}\right)
    -  \left[ \: 
             \frac{L_v}{T} \, + c_{pd} \, \gamma
         \: \right]  q_l
 \: -  \left[ \: 
             \frac{L_s}{T} \, + c_{pd} \, \gamma
         \: \right]  q_i
 \: + \: c_{pd} 
    \left[ \: 
     \Lambda - \gamma \: 
          \ln\!{\left( \frac{r_v}{r_{\ast}} \right)}^{\!}
    \: \right]  q_t
\label{eq_s2_plot} \: .
\end{align} 
The dry-air entropy reference value  $s_{d}^0$ is a constant which has no physical impact on computations of derivatives, gradients or comparisons of $s_{dry}$, $s_{abs}$, $s_{abs/1}$ or $s_{abs/2}$ between two points.

The second order correction terms are small in Equation~(\ref{eq_s2_plot})  because 
$L_v/T \approx 2.5\:10^6 / 280 \approx 8900$~J~K${}^{-1}$~kg${}^{-1}$ 
and 
$L_s/T \approx 2.8\:10^6 / 280 \approx 10000$~J~K${}^{-1}$~kg${}^{-1}$ 
are much larger than
$c_{pd} \, \gamma \approx 460$~J~K${}^{-1}$~kg${}^{-1}$.
Similarly, $\Lambda \approx 6 $ is larger than the term
$- \gamma \: \ln( r_v / r_{\ast} )$ because this term exactly cancels out for $r_v \approx r_{\ast} = 12.4$~g~kg${}^{-1}$ and is smaller than $ \pm \gamma \approx \pm 0.46$ for $ 4.6 < r_v < 33.7 $~kg${}^{-1}$.
This term becomes larger for smaller values of $r_v$, but the risk of infinite values for very small values of $r_v$ does not exist, because the product $\ln(r_v) \: q_v$ has the limit $0$ when $r_v \rightarrow 0$ and $q_v \rightarrow 0$ (see M11 \citep{Marquet_2011}).






The numerical evaluations and meteorological studies of $s_{abs}$ and/or $\theta_s$ were carried out in a series of a papers published since 2011:
the discovery of the well-mixed character for absolute entropy in PBL 
of marine stratocumulus \cite{Marquet_2011};
Brunt-V\"{a}is\"{a}l\"{a} frequency calculations 
of moist air \cite{Marquet_Geleyn_2013};
the computation and plot of the (absolute) potential vorticity 
$PV(s_{abs})$ \cite{Marquet_2014};
a synthesis of previous works \cite{Marquet_Geleyn_2015};
a study of a simulated Huricane \cite{Marquet_2017};
computations of marine bulk exchange coefficients \cite{Marquet_Belamari_2017_WGNE_Lewis};
computation of atmospheric turbulent exchange coefficients
\cite{Marquet_al_2017_WGNE_Lewis};
and the study of Hector-the-Convector simulated with 
Meso-NH in Giga-LES mode \cite{Marquet_Dauhut_2018}.

Differently, most studies of atmospheric entropy are carried out with hypotheses for reference entropies that are different from those deduced from the third principle of thermodynamics.
It is commonly assumed that it is possible to cancel the reference entropies of dry air ($s_d^0 = 0$) and liquid water ($s_l^0 =0$) at a temperature of zero degrees Celcius ($273.15$~K). 
This method leads to the formulations studied in Emanuel's and Pauluis' approaches (E94 \citep{Emanuel94}, P11 \citep{Pauluis_2011}, \citep{Mrowiec_al_2016}), which are based on specific moist-air entropies that can be written as
\begin{align}
 s_{P11}  & \: = \: \left( s_{abs} \: - \:  s_{d}^0 \right)
    + \left( s_{d}^0 - s_{l}^0 \right) \: q_t
  \label{eq_s_P11}
  \: , \\
 s_{E94}  
     \: - \:
    \left[  
    c_{pd} \: \ln(T_0)
    \: - \:  
     R_d \: \ln(p_0) 
    \right]
 & \: = \: s_{P11} 
    \: + \:  
    \left[ R_d \: \ln(p_0) 
    + (c_l-c_{pd}) \: \ln(T_0) 
    \right]  \: q_t 
  \label{eq_s_E94}
  \: .
\end{align}
The constant terms $s_{d}^0$ and 
 $[ \: R_d \: \ln(p_0) \: - \: c_{pd} \: \ln(T_0) \: ]$
has no physical impact and the true differences between $s_{abs}$, $s_{P11}$ and $s_{E94}$ are the terms in the right-hand sides that depend on the total water $q_t$, which is variable in time and space.
Formulations E94 and P11 correspond to the use of the equivalent potential temperature $\theta_e$ which can be written, at the first order
\begin{align}
  {({\theta}_{e})}_1
   & = \: \theta \; \:
    \exp\! \left( \: 
       \frac{L_v \: q_v }{c_{pd} \: T} 
       \right)
\label{eq_thetae1} \: .
\end{align}

IAPWS-2010 and TEOS-10 formulations (\citep{Feistel_al_2010} \citep{Feistel_2018}) calculated with the SIA software 
(\url{http://www.teos-10.org/software.htm}) 
use the same assumptions $s_d^0 = $0 and $s_l^0 = $0 as in E94 and P11, and give about the same moist-air entropy as $s_{P11}$ given by Equation~(\ref{eq_s_P11}) (preliminary results are described in \url{https://arxiv.org/abs/1901.08108}).
    
The purpose of this study is not to demonstrate the benefits or realism of a particular formulation, an aspect that is controversial in the community where very few people believe in a possible impact of the third law formulations $s_{abs}$ on the physics of the atmosphere.
It is more simply to show that these terms in Equations~(\ref{eq_s_P11}) and (\ref{eq_s_E94}) that depend on $q_t$ may induce important differences for the evaluation of the surface fluxes of moist-air entropy, and therefore the necessary imbalance inducing the entropy production by the atmosphere in terms of $s_{abs}$, $s_{P11}$ or $s_{E94}$.
This study is also an opportunity to show that the entropy fluxes calculated with first and second order approximations $s_{abs/1}$ and $s_{abs/2}$ are very close to those calculated with $s_{abs}$.

\section{The surface budget of entropy for the turbulent atmosphere} 
\label{Entropy_surf_fluxes}

In most studies dealing with moist processes (\cite{Peixoto_al_1991} \cite{Stephens_OBrien_1993} \cite{Goody_2000} \cite{Pauluis_Held_2002_I} \cite{Ozawa_2003} \cite{Kleidon_Lorenz_2005} \cite{Fraedrich_Lunkeit_2008} \cite{Kleidon_2010} \cite{Pauluis_al_2010} \cite{Pauluis_2011} \cite{Pascale_al_2011} \cite{Lucarini_al_2014}), the entropy balance equation is calculated with Equation~(\ref{Eq_Entropy_1}), where the impact of the sensible (thermal) and latent (water) heat fluxes $SH$ and $LH$ are estimated via the flux $\vec{J}_s$ by adding the terms $SH/T$ and $LH/T$ as in  Equation~(\ref{Eq_Entropy_1_cond}), where $T$ is the surface temperature.

However, sensible and latent heat fluxes are not defined for the atmosphere at the laboratory scale in GCMs and PNT models (e. g. from a few centimetres to a few metres), but rather as turbulent fluxes for grid-cells of several kilometres long and via Reynolds axioms.
Such a separation of the entropy equation in mean terms plus associated turbulent flows was made by \citeauthor{Herbert_3_1975} \cite[Equations~III-3 to III-5]{Herbert_3_1975} and similar calculations are made in \citeauthor{Gassmann_2018}'s papers (\cite{Gassmann_al_2015} \cite{Gassmann_2018}) and in \cite{rogachevskii_kleeorin_2015} through the average version of Equation~(\ref{Eq_Entropy_1}):
\begin{align} 
 \frac{\partial (\, \overline{\rho} \, \overline{s} \,)}{\partial t}
 + \,
 \vec{\nabla} . 
 \left( 
   \overline{\rho} \, \overline{s} \, \overline{\vec{u}} 
 \right)
 \: = \:
 \overline{\rho} \:
       \left[
 \frac{\partial \, \overline{s}}{\partial t}
 + \,
 \left( \overline{\vec{u}} . \vec{\nabla} \right)
 \overline{s}
        \right]
 \: = \:
 - \,
 \vec{\nabla} . 
 \left( \overline{\rho} \, \overline{ s \, \vec{u}} \right)
 \, - \, \overline{\vec{\nabla} . \vec{J}_R}
 \, - \, \overline{\vec{\nabla} . \vec{J}_s}
 \, + \, 
 \overline{\sigma} \: .
 \label{eq_s_turb} 
\end{align} 
The turbulent fluxes are generated by the term
$- \,\vec{\nabla} .
\left( 
  \overline{\rho} \, \overline{ s \, \vec{u}} 
\right)
$,
rather than by diabatic terms included in
$- \, \overline{\vec{\nabla} . \vec{J}_s}$ via molecular fluxes$\vec{J}_s$  associated with $\overline{\sigma}$ according to Equation~(\ref{Eq_Entropy_1_cond}).
This is one of the important points that differentiates this study and few others 
(\cite{Herbert_3_1975},
\cite{Gassmann_al_2015},
\cite{rogachevskii_kleeorin_2015},
\cite{Gassmann_2018}) from most other approaches, where fluxes of entropy are not considered in this way.
 

Only the vertical part of turbulent fluxes
$- \,\vec{\nabla} .
\left( 
  \overline{\rho} \, \overline{ s \, \vec{u}} 
\right)
$
are involved in GCM and PNT models. 
According to \cite{rogachevskii_kleeorin_2015} these vertical turbulent fluxes must be calculated according to 
$- \: \partial/\partial z \, ( FS )$ with
\begin{align}
FS & = \: \overline{\rho} \; \overline{w' s'} \: ,
\end{align}
rather than via with the alternative formula
$- \: (1/\overline{T})\: 
\partial/\partial z \, 
( \overline{\rho} \: \overline{T} \: \overline{w' s'})$.

The study of surface entropy fluxes can be simplified by assuming from here that there is no condensed water close to the ground ($q_l=q_i=0$), with sensible and latent heat fluxes defined by the usual formulations used in GCM and NWP models:
\begin{align}
SH & = \overline{\rho} \; c_{pd} \: \overline{w' \theta'} 
 \label{eq_SH} 
\: , \\
LH & = \overline{\rho} \; L_v \: \overline{w' q'_v} 
 \label{eq_LH} 
\: .
\end{align}

The turbulent fluxes of entropy calculated from Equations~(\ref{eq_s_thetas_plot}), (\ref{eq_thetas}), (\ref{eq_s1_plot}) and (\ref{eq_s2_plot}) can be written as a weighted sum of $SH/T$ and $LH/T$, leading to
\begin{align}
FS & = \,
 Y_{SH} \: \left( \frac{SH}{T}  \right)
  \: + \: 
 Y_{LH} \:\left( \frac{LH}{T}  \right) \: .
 \label{eq_FS} 
\end{align}

If the terms $Y_{SH}$ and $Y_{LH}$ are equal to $1$ in current studies of the surface budget of atmospheric entropy, the first order coefficients corresponding to ${({\theta}_{s})}_1$ are written
\begin{align}
Y_{SH/1} & = \:
\frac{T}{\theta}
\:= \:
\left(\frac{p}{p_0}\right)^{\kappa}
 \label{eq_FSH1} 
\: , \\
Y_{LH/1} & = \: 
    \left(\frac{c_{pd}\: T}{L_v}\right)\: \Lambda
 \label{eq_FLH1} 
\: ,
\end{align}
where the factor ${T}/{\theta}$ is close to $1$ in the plain (where $p \approx p_0$) for $SH/T$, but where the factor 
${c_{pd}\: T \: \Lambda \,}/{\,L_v}
\approx 1005 \times 290 \times 6 \: / \: 2.5\:10^6
\approx 0.7$ is always different from unity for $LH/T$.
This factor, which is $30$~\% lower than in current studies, must modify the value of the surface entropy flux quite strongly.
A rough estimate can be obtained with the standard values 
$SH \approx 15$~W~m${}^{-2}$ and $LH \approx 80$~W~m${}^{-2}$, 
giving $(SH+LH)/T \approx 0.328$~W~m${}^{-2}$~K${}^{-1}$
and $(SH+0.7\,LH)/T \approx 0.245$~W~m${}^{-2}$~K${}^{-1}$.
This corresponds to a $25$~\% reduction in total entropy flux and a long-term and global average difference of $0.083$~W~m${}^{-2}$~K${}^{-1}$, which can be important in entropy budget calculations.

Second order coefficients are written as
\begin{align}
Y_{SH/2} & = \:
\frac{T}{\theta}
 \label{eq_FSH2} 
\: , \\
Y_{LH/2} & = \: \left(\frac{c_{pd}\: T}{L_v}\right) \: 
 \left[ \: 
 \Lambda
 \: - \: \frac{\gamma}{1-q_v}
 \: - \: \gamma \: 
 \ln{\left( \frac{r_v}{r_{\ast}} \right)}^{\!}
 \: \right]
 \label{eq_FLH2} 
\: ,
\end{align}
while the most accurate versions are written
\begin{align}
Y_{SH} & = \:
\frac{T}{\theta} \: (1 + \lambda \: q_v)
 \label{eq_FSH} 
\\
Y_{LH} & = \: \left(\frac{c_{pd}\: T}{L_v}\right) \: 
 \left[ \: 
 \Lambda
 \: - \: \frac{\gamma}{1-q_v}
 \: - \: \gamma \: 
     \ln{\left( \frac{r_v}{r_0} \right)}^{\!}
 \: - \: \kappa \: \delta \: 
     \ln{\left( \frac{p}{p_0} \right)}
 \: - \: \lambda \:  
     \ln{\left( \frac{T}{T_0} \right)}
 \: \right. 
 \nonumber \\
       & \;\;\;\;\;\;\;\;\;\;\;\;\;\;\;\;\;\;\;\;\;\;\;
 \left. 
   + \: \gamma 
     \: \left(\frac{1+\delta \: q_v}{1+\eta \: r_v}\right)
     \: \left(\frac{1 + r_v}{1-q_v}\right)
   + \: \kappa \: \delta \: 
     \ln\left( \frac{1 + \eta \: r_v}{1 + \eta \: r_0} \right)
 \: \right] \: .
 \label{eq_FLH} 
\end{align}

The impact of corrective terms must be small compared to those of first-order formulations (for instance 
${\gamma}/{(1-q_v)} \approx 0.46 \ll \Lambda \approx 6$).
This must be verified from model outputs and observations in the following sections.

\section{Results} 

\label{Results}

\subsection{Entropy structures and fluxes for ARPEGE analyses} 
\label{Results_ARPEGE}

\begin{figure}[H]
\centering
\includegraphics[width=7.7 cm]{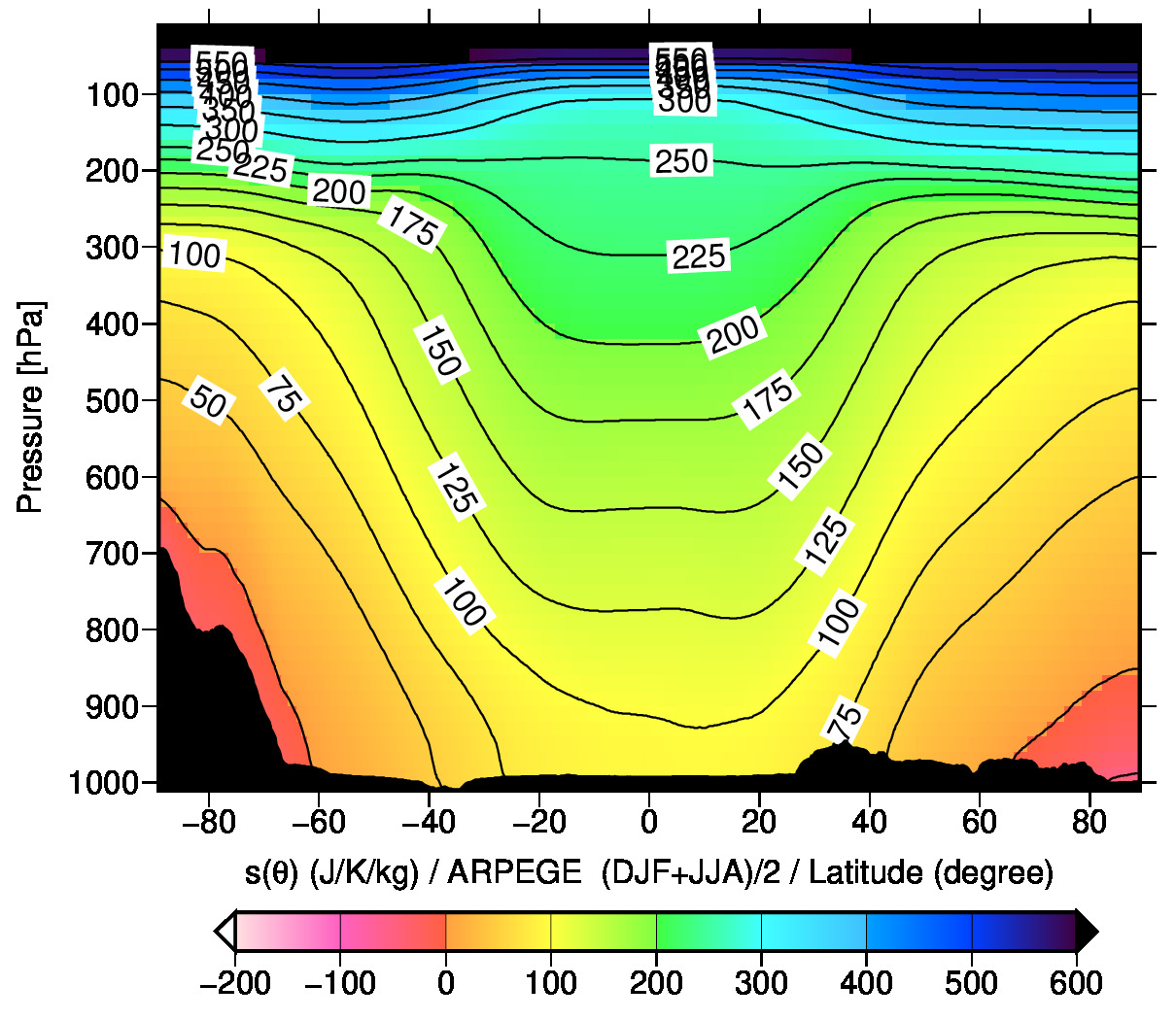}
\includegraphics[width=7.7 cm]{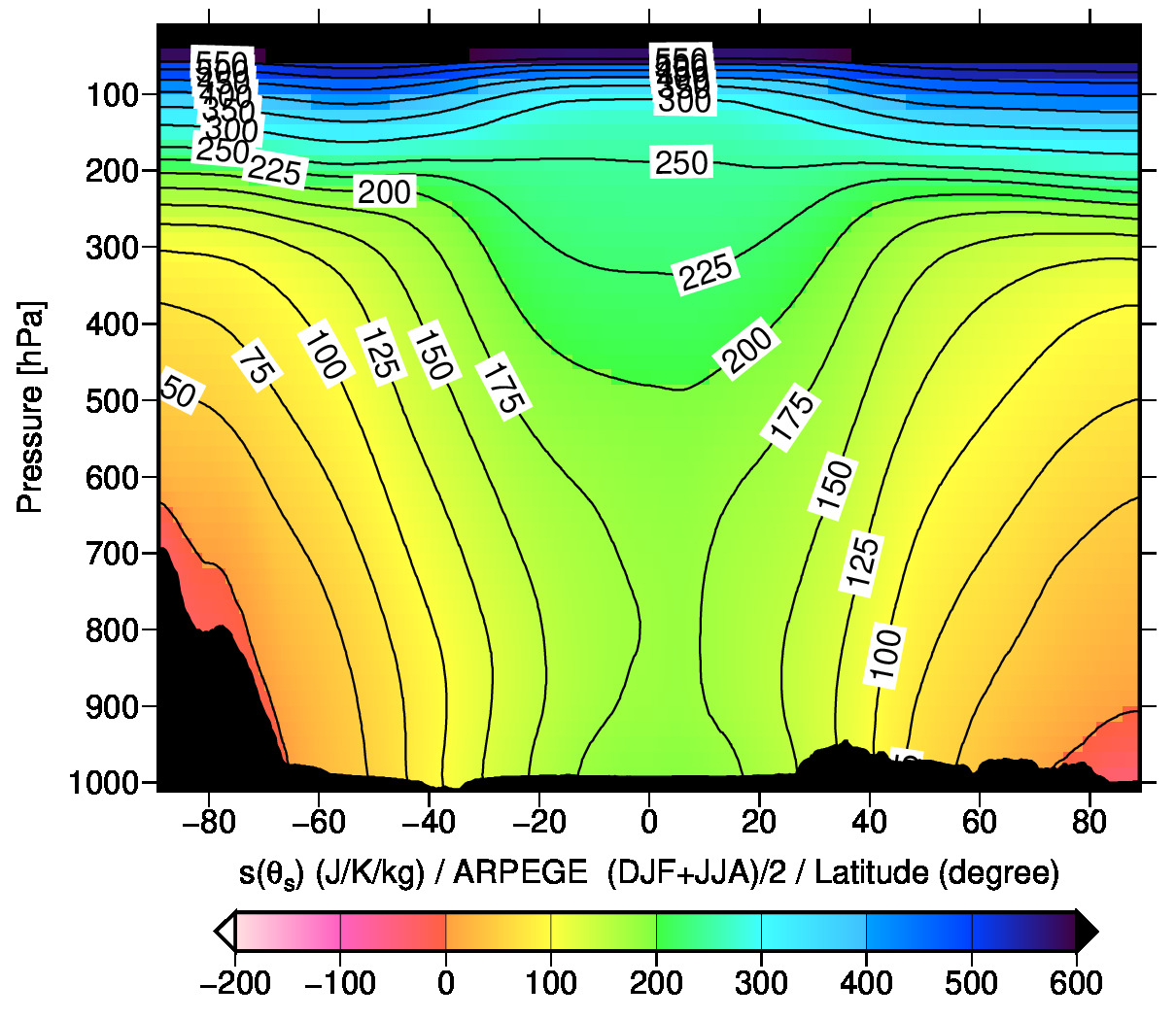} \\
\vspace*{-1.5mm}
(\textbf{a}) \hspace*{7 cm} (\textbf{b}) \\
\includegraphics[width=7.7 cm]{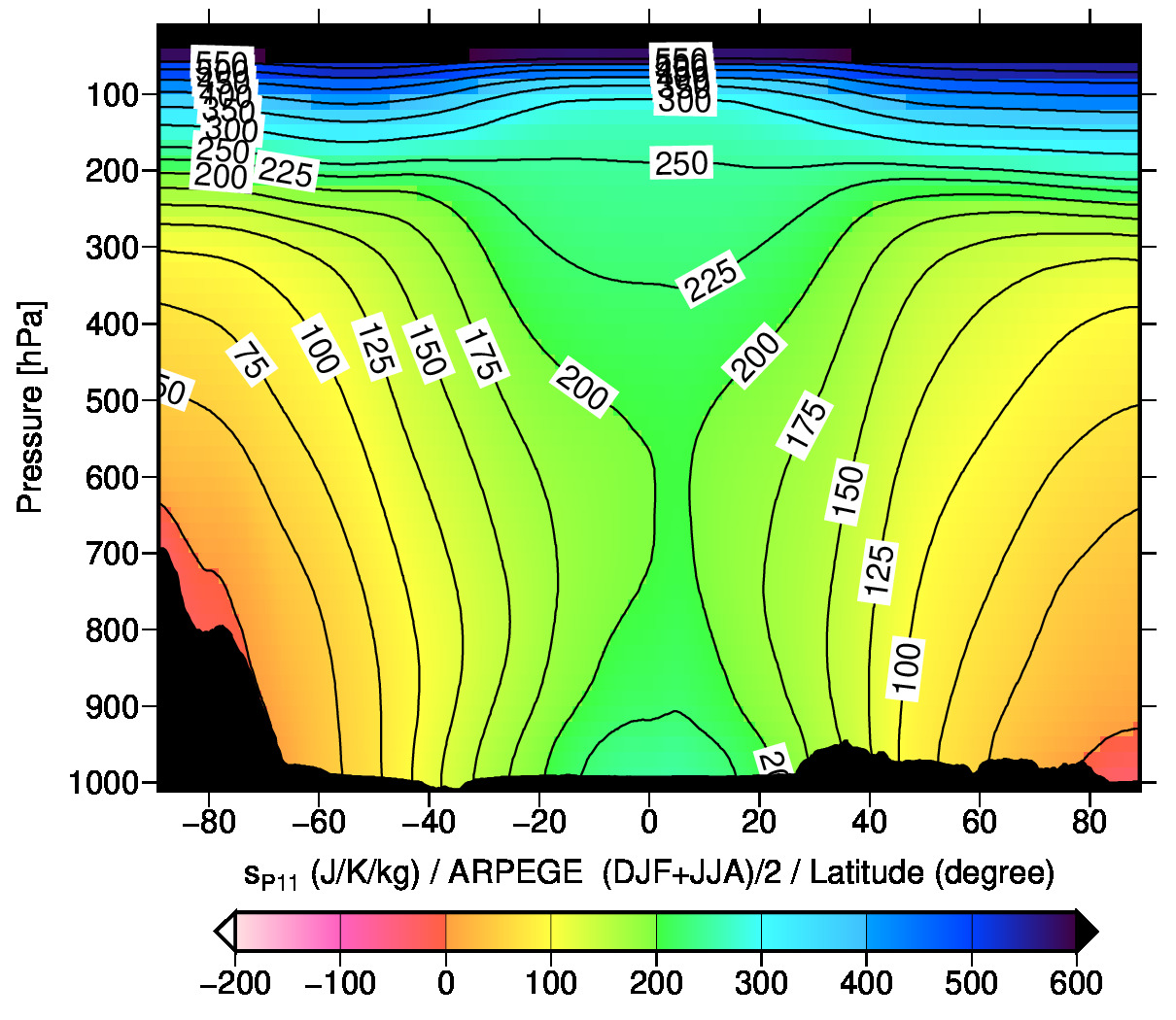}
\includegraphics[width=7.7 cm]{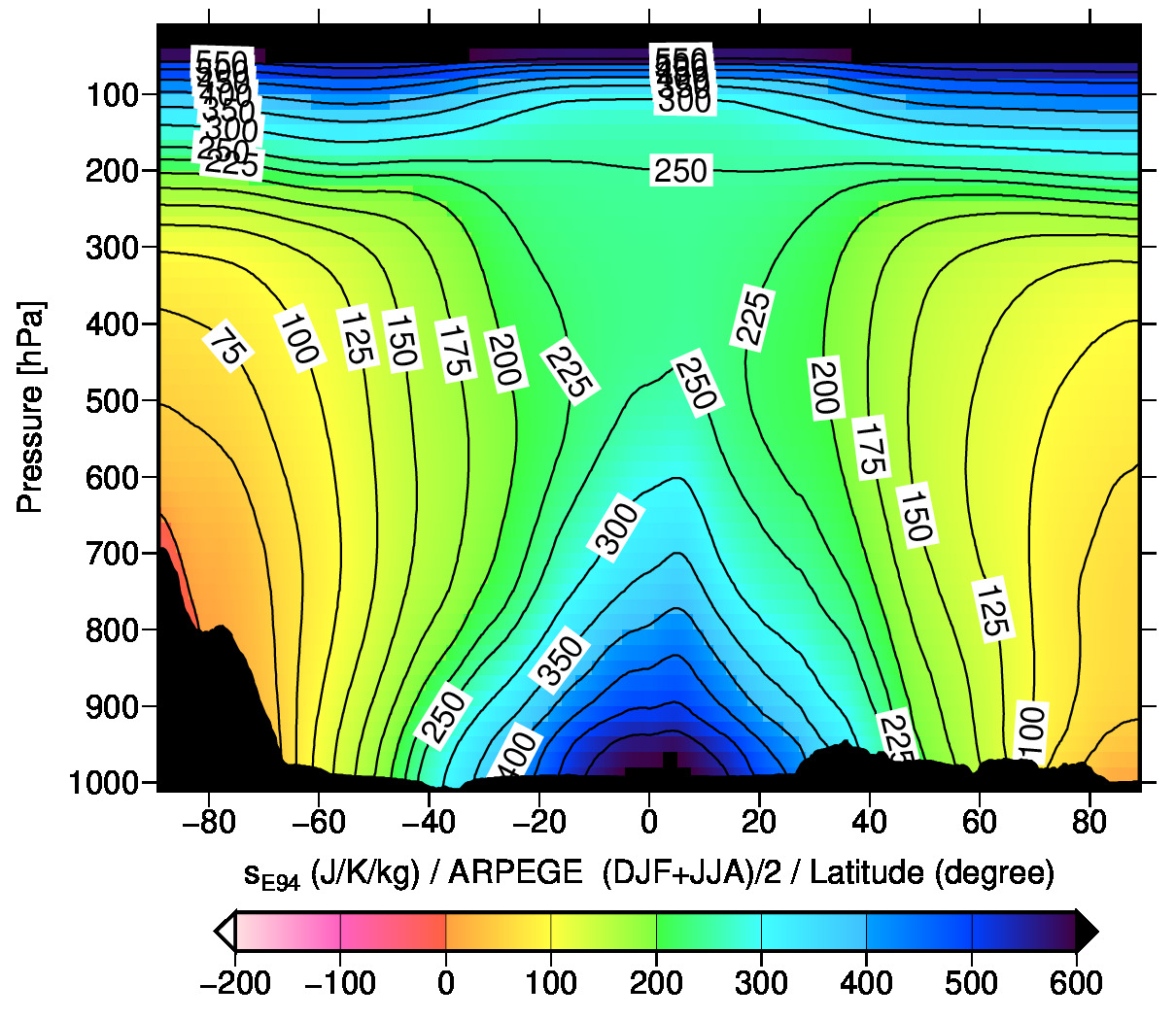} \\
\vspace*{-1.5mm}
(\textbf{c}) \hspace*{7 cm} (\textbf{d})
\caption{The zonal and 6-months averages (DJF+JJA)/2 of entropies for ARPEGE analyses.
(\textbf{a}): the dry air version with $s_{dry} = s(\theta)$;
(\textbf{b}): the absolute (third law) values $s_{abs}$ (HH87, M11, SS19) with $\theta_s$;
(\textbf{c}): the P11 value $s_{P11}$  with $\theta_e$;
(\textbf{d}): the E94 value $s_{E94}$ with $\theta_e$.
The spacings are $25$ units up to $250$~J~K${}^{-1}$~kg${}^{-1}$, with spacings of $50$ units above the troposphere.
}
\label{Figure_ARPEGE_mean}
\end{figure}

The Arpege NWP global system \cite{Courtier_al_1991} \cite{Courtier_al_1994} \cite{Auger_al_WGNE_2016} 
has a variable horizontal resolution grid of about 10 km over Western Europe that increases up to about 30 km close to Australia.
The analyses files available every 6 hours at the  horizontal resolution of $0.25$~degree were averaged for December 2018 and January to August 2019.
Figure~\ref{Figure_ARPEGE_mean} shows the zonal and 6-months averages for the dry- and moist-air entropies:
$s_{dry}(\theta)-s_d^0$; 
$s_{abs}(\theta_s)-\:s_d^0$;
$s_{P11}(\theta_e)$; and
$s_{E94}(\theta_e)-c_{pd}\:\ln(T_0/p_0^\kappa)$ 
given by 
Equations~(\ref{eq_s_theta_dry_plot}),
(\ref{eq_s_thetas_plot}),
(\ref{eq_s_P11}) and 
(\ref{eq_s_E94}).

The lines of equal values of $s_{dry}$ correspond to the usual climatological averages in (a), with small polar values and high equatorial values in the troposphere.

The lines of equal values of absolute entropy $s_{abs}$ are different in (b), especially in the lower tropical troposphere, where the specific humidity $q_t$ is the larger.
The values of $s_{abs}$ are almost vertical in the lower tropical troposphere and in the mid-latitudes.
This must correspond to the well-mixed entropy character that has been observed in most of marine boundary layers \cite{Marquet_2011}.
The slopes of $s_{dry}$ and $s_{abs}$ are similar in the mid-troposphere and for mid-latitudes.
This is an important aspect that preserves the vision of the dynamics that follow the slopes of dry or moist-isentropes, with differences that would be interesting to examine in more detail in a future study.

The structure in (c) for $s_{P11}$ shows larger impacts of $q_t$, with less vertical isolines and larger vertical gradients in the lower tropical layers, where an area of high values is beginning to appear.
These aspects seem to be accentuated with $s_{E94}$, with larger values in the whole lower troposphere, marked vertical gradients and a clear area of high values within the tropics.

\begin{figure}[H]
\centering
\includegraphics[width=7.7 cm]{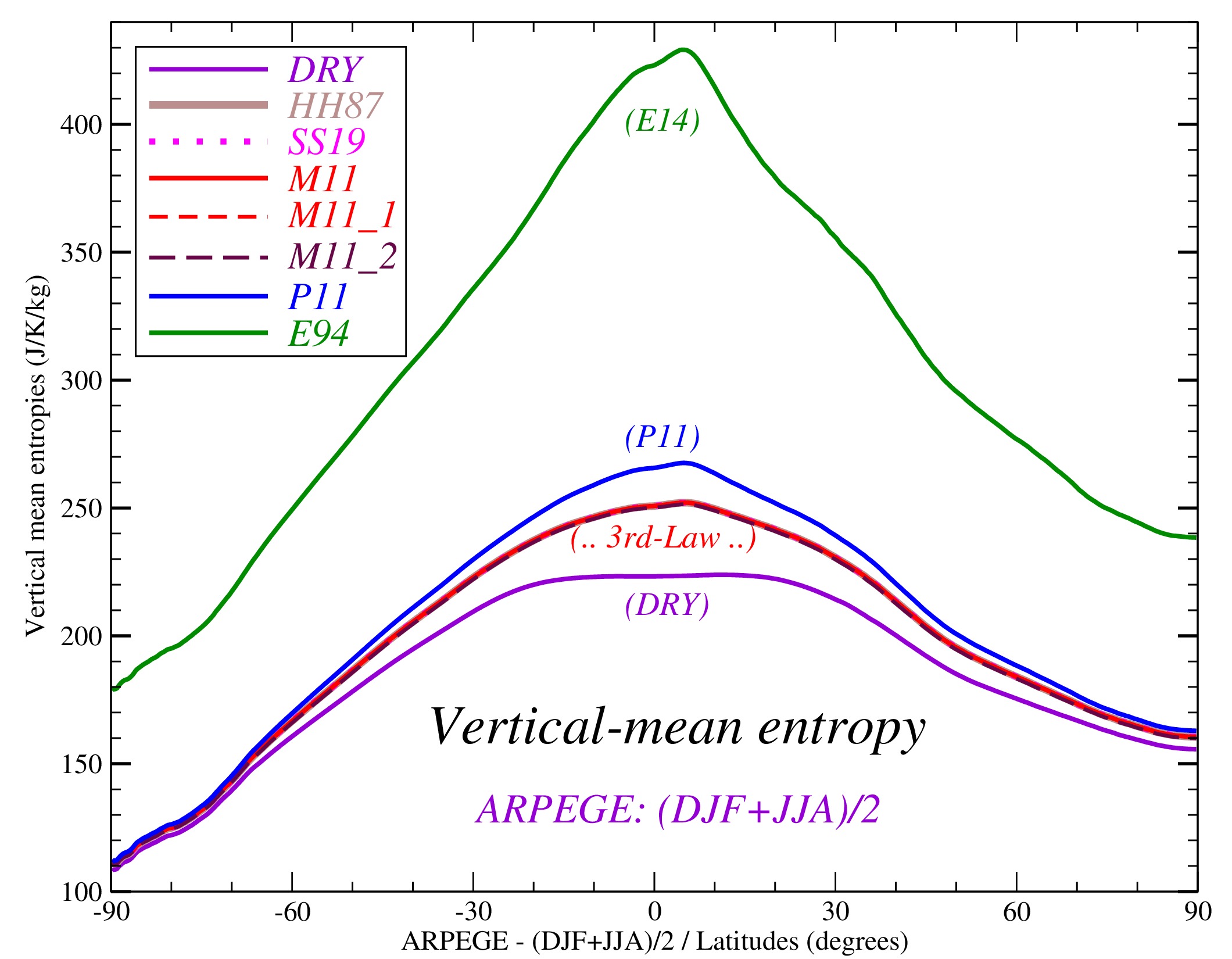}
\includegraphics[width=7.7 cm]{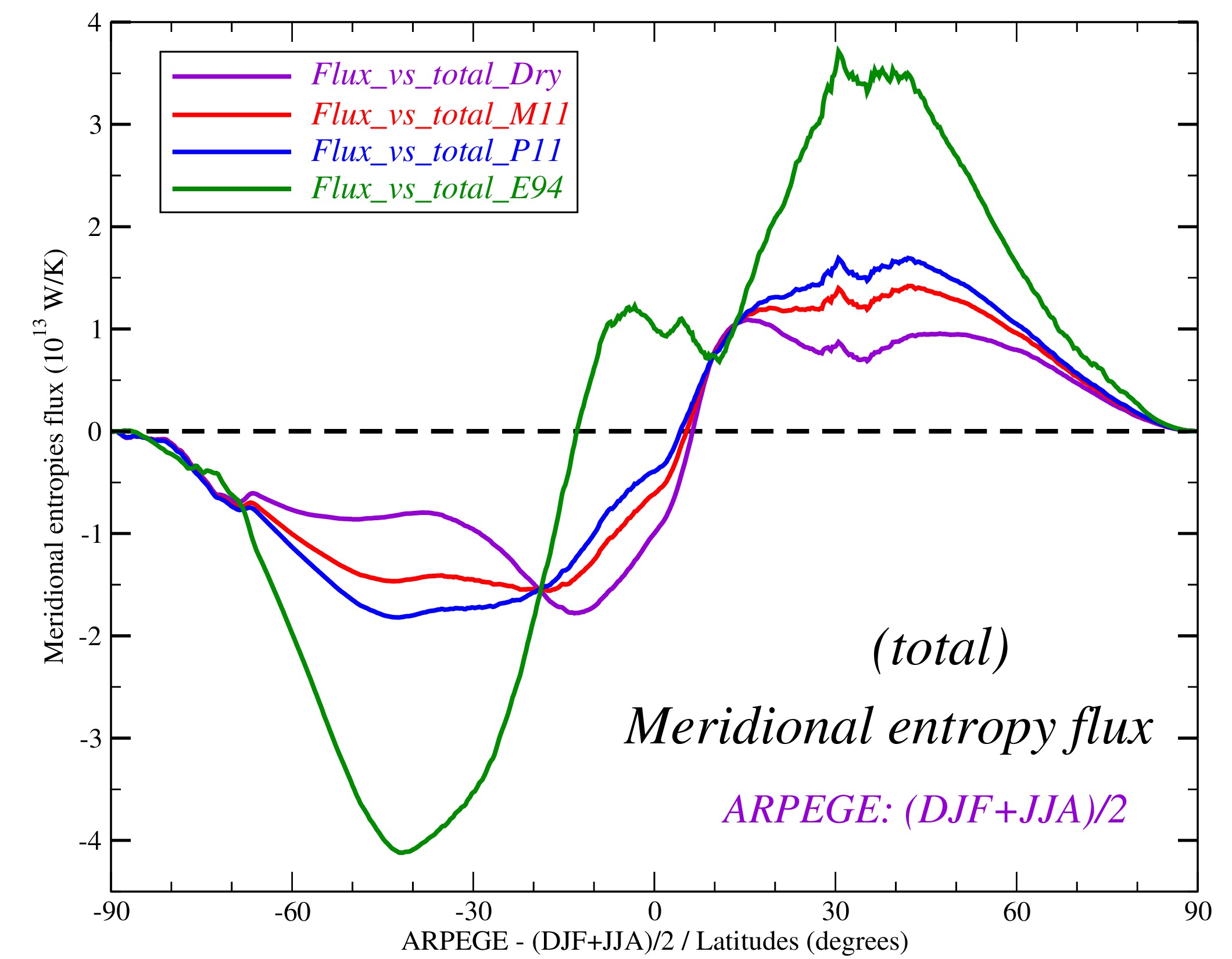}\\
\vspace*{-1.mm}
(\textbf{a}) \hspace*{7 cm} (\textbf{b}) \\
\includegraphics[width=7.7 cm]{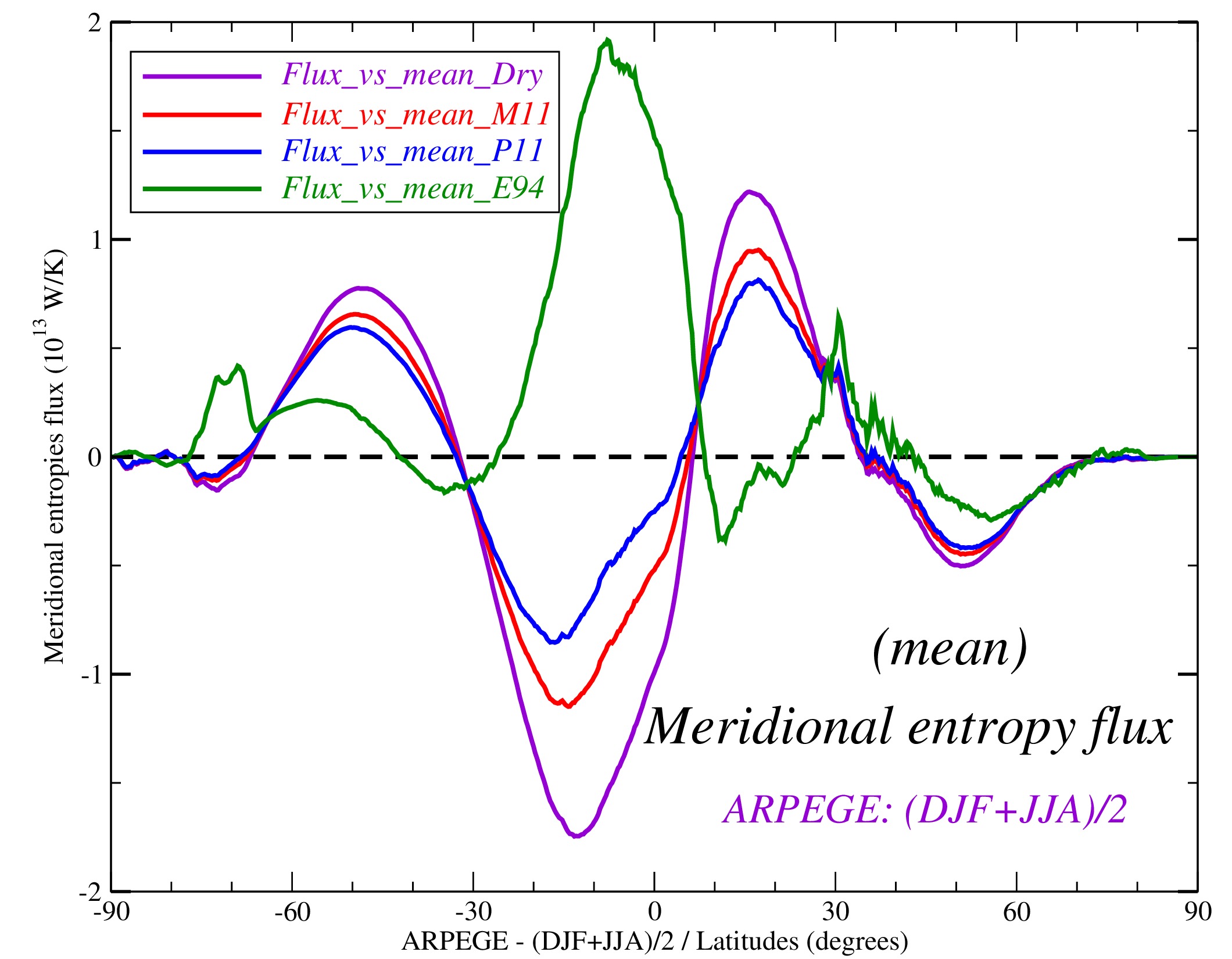}
\includegraphics[width=7.7 cm]{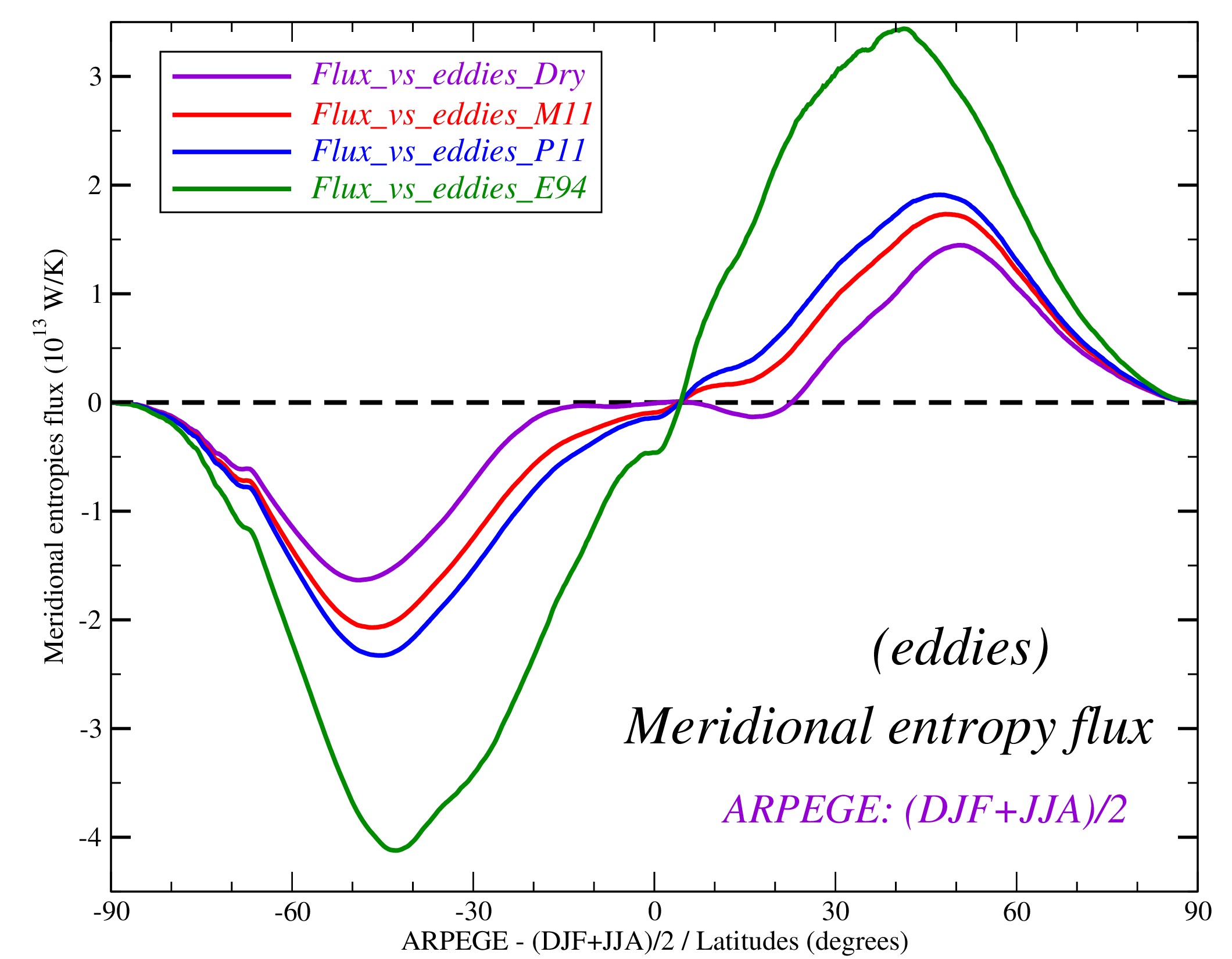}\\
\vspace*{-1.mm}
(\textbf{c}) \hspace*{7 cm} (\textbf{d}) \\
\caption{Mean entropy and entropy fluxes ($>0$ northward) computed with the same ARPEGE dataset as in Figure~\ref{Figure_ARPEGE_mean}.
(\textbf{a}): vertical-mean entropies (J~K${}^{-1}$~kg${}^{-1}$).
(\textbf{b}): poleward flux of entropies ($10^{13} \,$~W~kg${}^{-1}$) for the total circulation, in (\textbf{c}) for the mean circulation and in (\textbf{d}) for the eddies.
}
\label{Figure_ARPEGE_vertical_mean_s}
\end{figure}

The translation of these differences is shown in 
Figure~\ref{Figure_ARPEGE_vertical_mean_s}~(a)
for the values of entropies integrated over the entire vertical. 
There are 4 new curves.
Those for $s_{abs/HHH87}-\:s_d^0$ and $s_{abs/SS19}-\:s_d^0$
correspond to articles \cite{Hauf_Holler_1987} and \cite{Stevens_Siebesma_2019} where the same third law standard values of entropies are used as in \cite{Marquet_2011}.
Those for $s_{abs/1}-\:s_d^0$ and $s_{abs/2}-\:s_d^0$
have been calculated with the equations~(\ref{eq_s1_plot})
and (\ref{eq_s2_plot}) for first- and second-order approximations of $\theta_s$.

Clearly, all the entropy formulations based on the third law  coincide en (a). 
It is hardly possible to differentiate them and they are in position $2/3$ between $s_{dry}$ and $s_{P11}$.
This result is general (\cite{Marquet_2011} \cite{Marquet_2017}) and can be explained by the ratio of about $9/6=2/3$ between $\Lambda \approx 6$ in Equation~(\ref{eq_thetas1}) for ${({\theta}_s)}_1$ and $L_v / (c_{pd} \: T) \approx 9$ in Equation~(\ref{eq_thetae1}) for ${({\theta}_{e})}_1$.

The values for $s_{E94}$ are much higher, in accordance with the high values in Figure~\ref{Figure_ARPEGE_mean}~(d) in the lower tropical layers.
We observe here the consequences of the terms 
$R_d \: \ln(p_0) \: q_t$ and
$(c_l-c_{pd}) \: \ln(T_0) \: q_t$
in  Equation~(\ref{eq_s_E94}),
which are inactive for quantities expressed "{\em per unit mass of dry air\/}" (see Equation~4.5.10 for $s/(1-q_t)$ in E94 \cite{Emanuel94}), but which become very active when transformed into specific quantities (i.e. per unit mass of humid air) and integrated over the whole atmosphere.\vspace*{-1.mm}

The poleward entropy transports shown in Figure~\ref{Figure_ARPEGE_vertical_mean_s}~(b,c,d) are similar to those computed in \cite{Pauluis_al_2010}.
The total meridian entropy flux 
$Fvs_{tot} = Fvs_{mean} + Fvs_{eddy}$
is calculated for each latitude as in \cite{Liang_al_2018} for the poleward energy transport, by the longitudinal and pressure integrals
\begin{align}
Fvs_{mean} & = \: 
2 \, \pi \: R_{earth} \: \cos(\phi)
\int_0^{p_s} \: \widehat{v} \; \;\widehat{s} \;\: \frac{dp}{g}
\: ,  
 \label{eq_F_vs_eddy} \\
Fvs_{eddy} & = \: 
2 \, \pi \: R_{earth} \: \cos(\phi)
\int_0^{p_s} \: \widehat{v'' \, s''} \; \: \frac{dp}{g}
\: ,
 \label{eq_F_vs_mean}
\end{align}

where the radius of the earth is $R_{earth} \approx 6.37\,10^6$~m and the acceleration of gravity is $g\approx 9.81$~m~s${}^{-2}$.
The eddy terms 
$v'' = v - \widehat{v}$ and 
$s'' = s - \widehat{s}$ indicate departures from the zonal averages $\widehat{v}$ and $\widehat{s}$ over each individual longitude circles of length $2 \, \pi \: R_{earth} \: \cos(\phi)$.

\begin{figure}[H]
\centering
\includegraphics[width=7.7 cm]{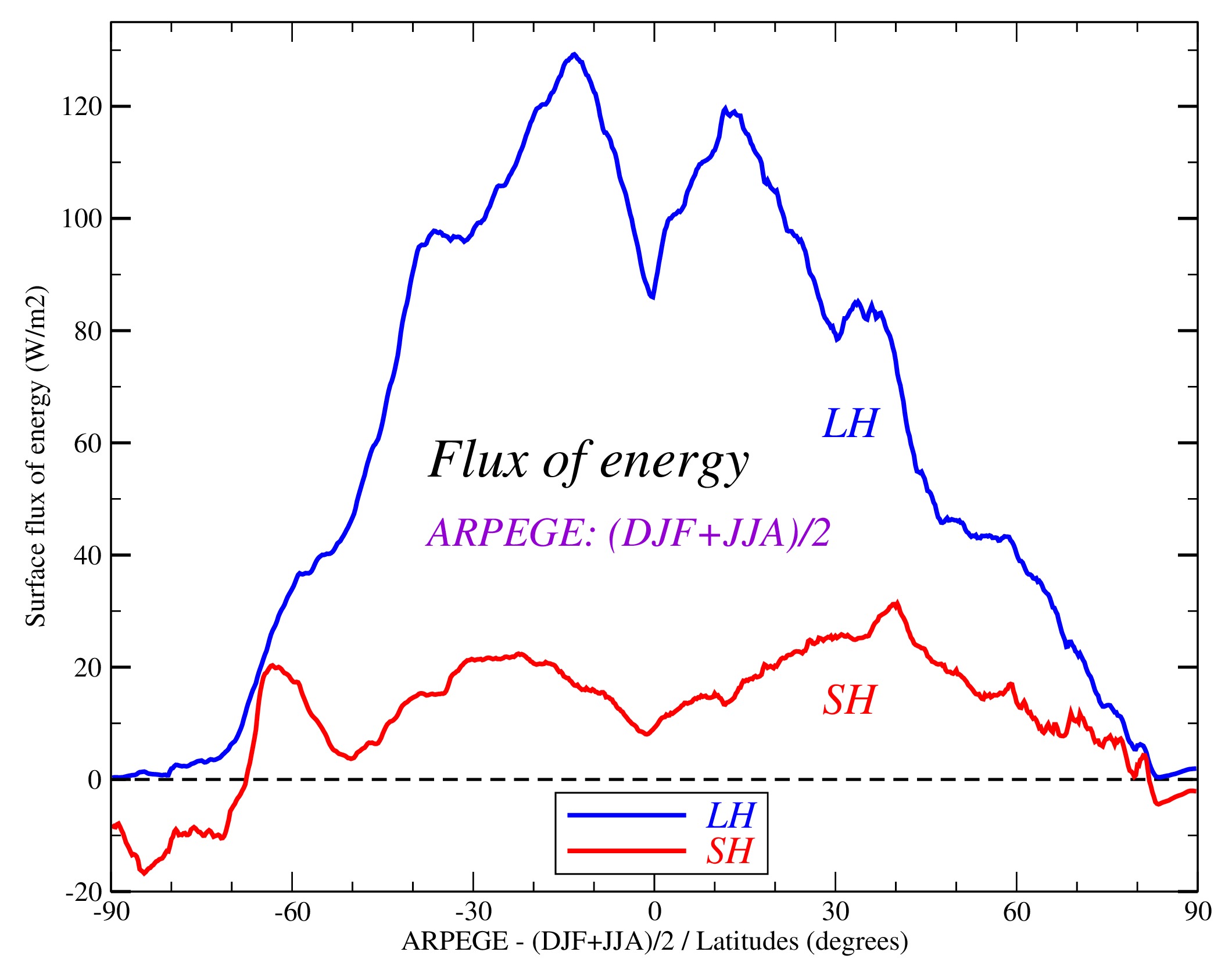}
\includegraphics[width=7.7 cm]{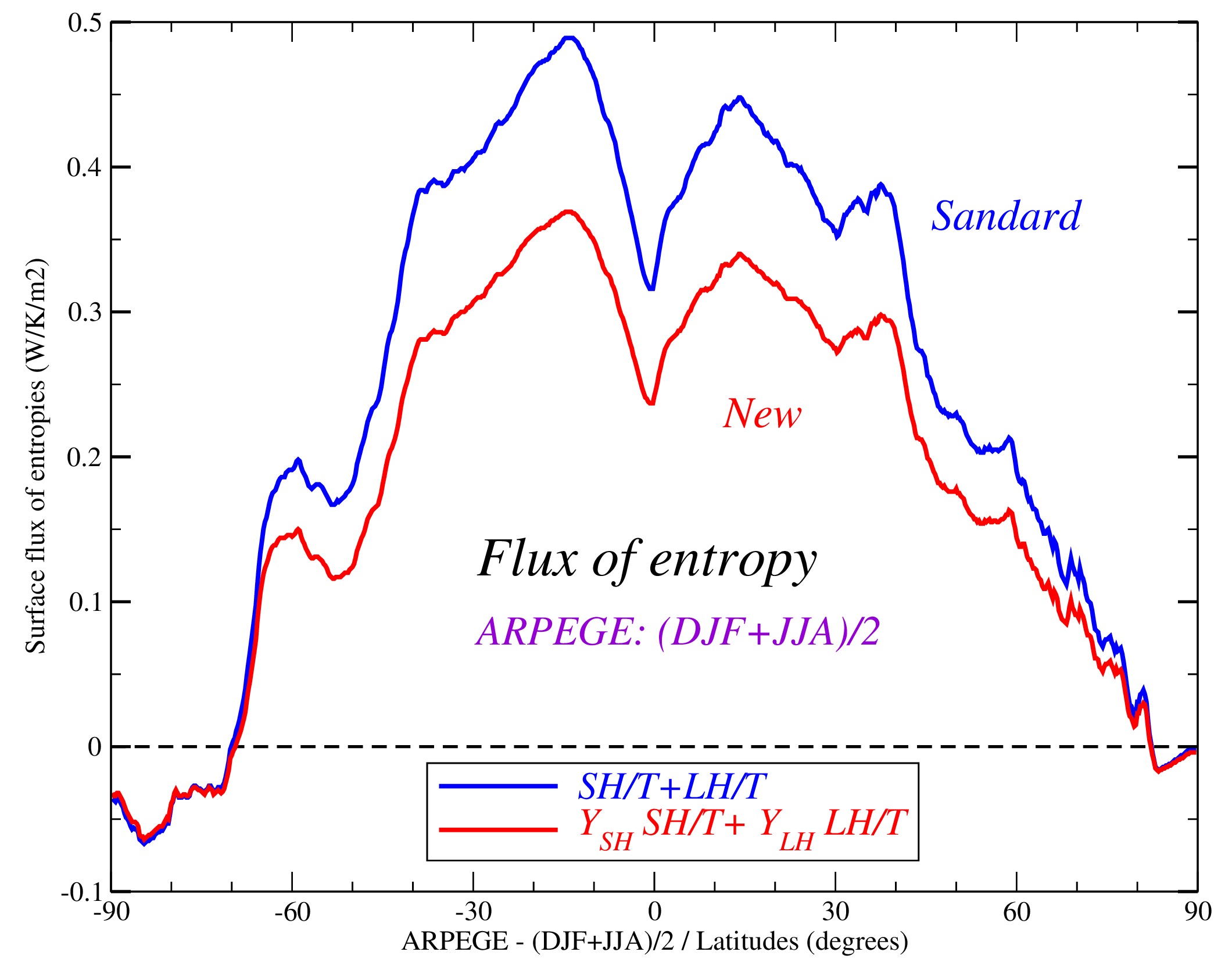} \\
\vspace*{-0.5mm}
(\textbf{a}) \hspace*{7 cm} (\textbf{b})
\caption{Surface fluxes ($>0$ upward) computed with the 
same ARPEGE dataset as in Figure~\ref{Figure_ARPEGE_mean}. 
(\textbf{a}) The sensible and latent surface fluxes of energy $SH$ and $LH$ 
   (W~m${}^{-2}$).
(\textbf{b}) The standard surface fluxes of entropy 
$SH/T+LH/T$
compared to the new version 
$Y_{SH} \: (SH/T) + Y_{LH} \: (LH/T)$ 
(W~K${}^{-1}$~m${}^{-2}$). 
}
\label{Figure_ARPEGE_LH_SH_all}
\end{figure}

Similar to the results in Figure~\ref{Figure_ARPEGE_vertical_mean_s}~(a) for the vertical average of entropies, the poleward entropy fluxes in (b,c,d) shows that the fluxes for the absolute (M11) version $s_{abs}(\theta_s)$ are intermediate and in a $2/3$ position between those for the dry air and P11 versions $s_{dry}(\theta)$ and  $s_{P11}(\theta_e)$.  
The poleward entropy flux for $s_{PE94}(\theta_e)$ are more different, especially in (c) for the transport by the average wind $\widehat{v}$.
This uncertainty regarding the transport by $\widehat{v}$ is the same as that described in  \cite{Liang_al_2018} for energy, with a large impact of the choice of integration constants and other constant terms such as 
$s_d^0$ and $c_{pd} \: \ln(T_0) - R_d \: \ln(p_0)$ 
in the left-hand side of Equations~(\ref{eq_s_theta_dry_plot}) to (\ref{eq_s_E94}).

However, the poleward transport in (d) generated by the eddies is not affected by these uncertainties, and the same differences between the $s_{dry}$, $s_{abs}$ (M11) and $s_{P11}$ formulations exist.
Moreover, negative values for $s_{dry}$ are noted between latitudes $5$ and $20$~North, while transports for $s_{abs}$ and $s_{P11}$ are both positive.
As for the differences between $s_{abs}$ and $s_{P11}$, the poleward transports evaluated with $s_{P11}$ systematically overvalue those calculated with $s_{abs}$.
Such differences are expected to have local and global impacts on entropy transport in the atmosphere.

\begin{figure}[H]
\centering
\includegraphics[width=7.7 cm]{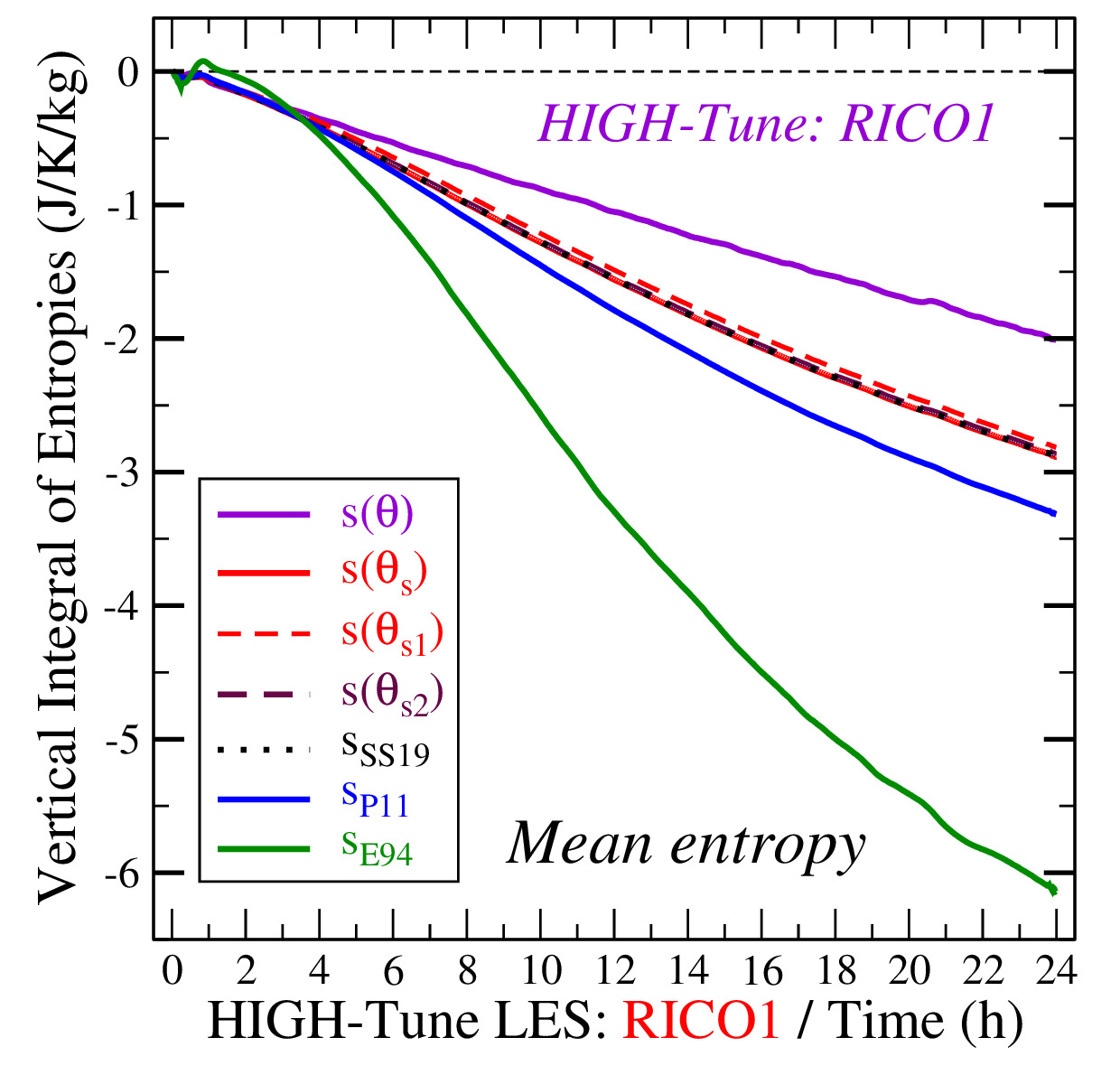}
\includegraphics[width=7.7 cm]{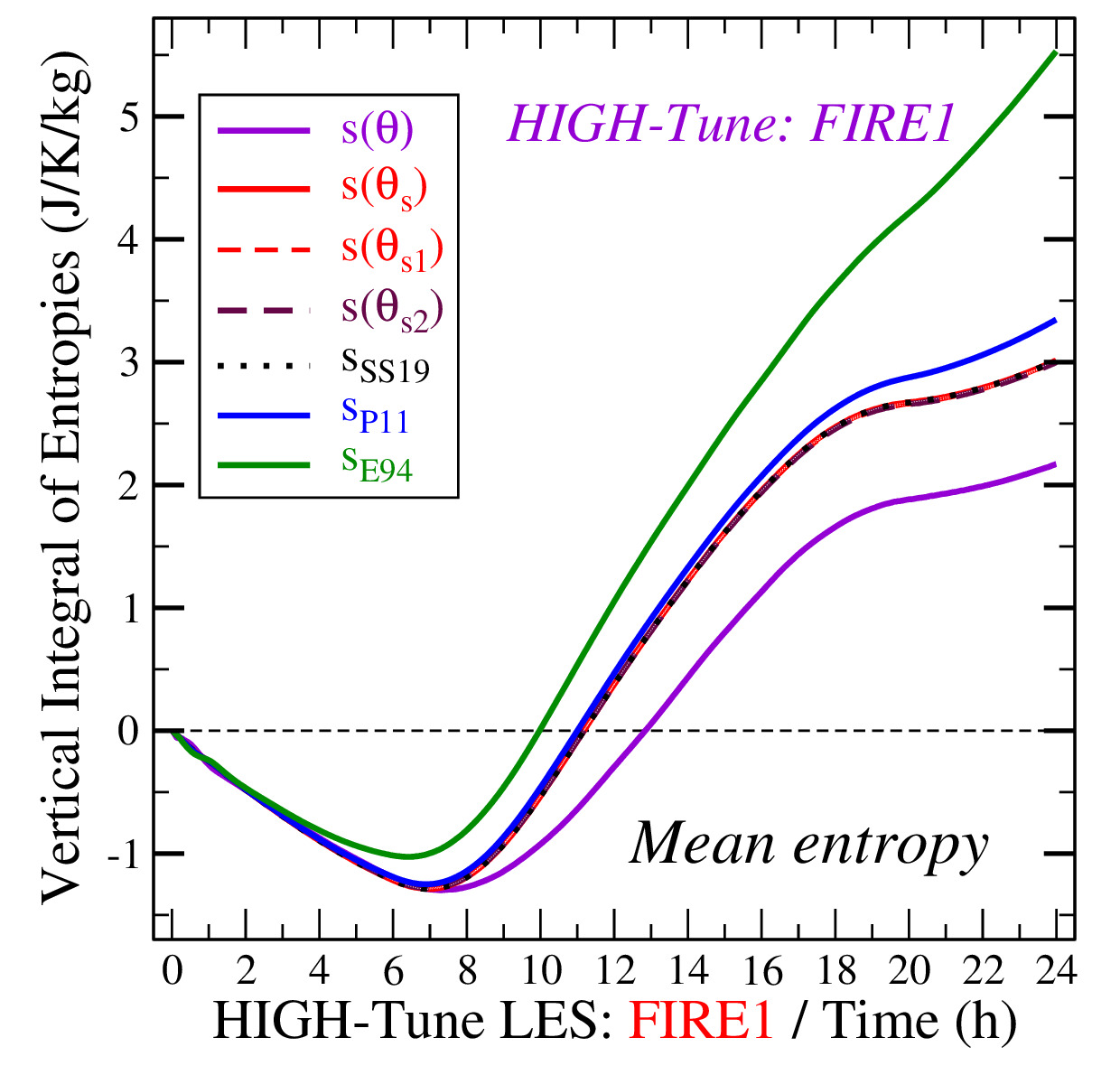} \\
\vspace*{-1.5mm}
(\textbf{a}) \hspace*{7.3 cm} (\textbf{b}) \\
\includegraphics[width=7.7 cm]{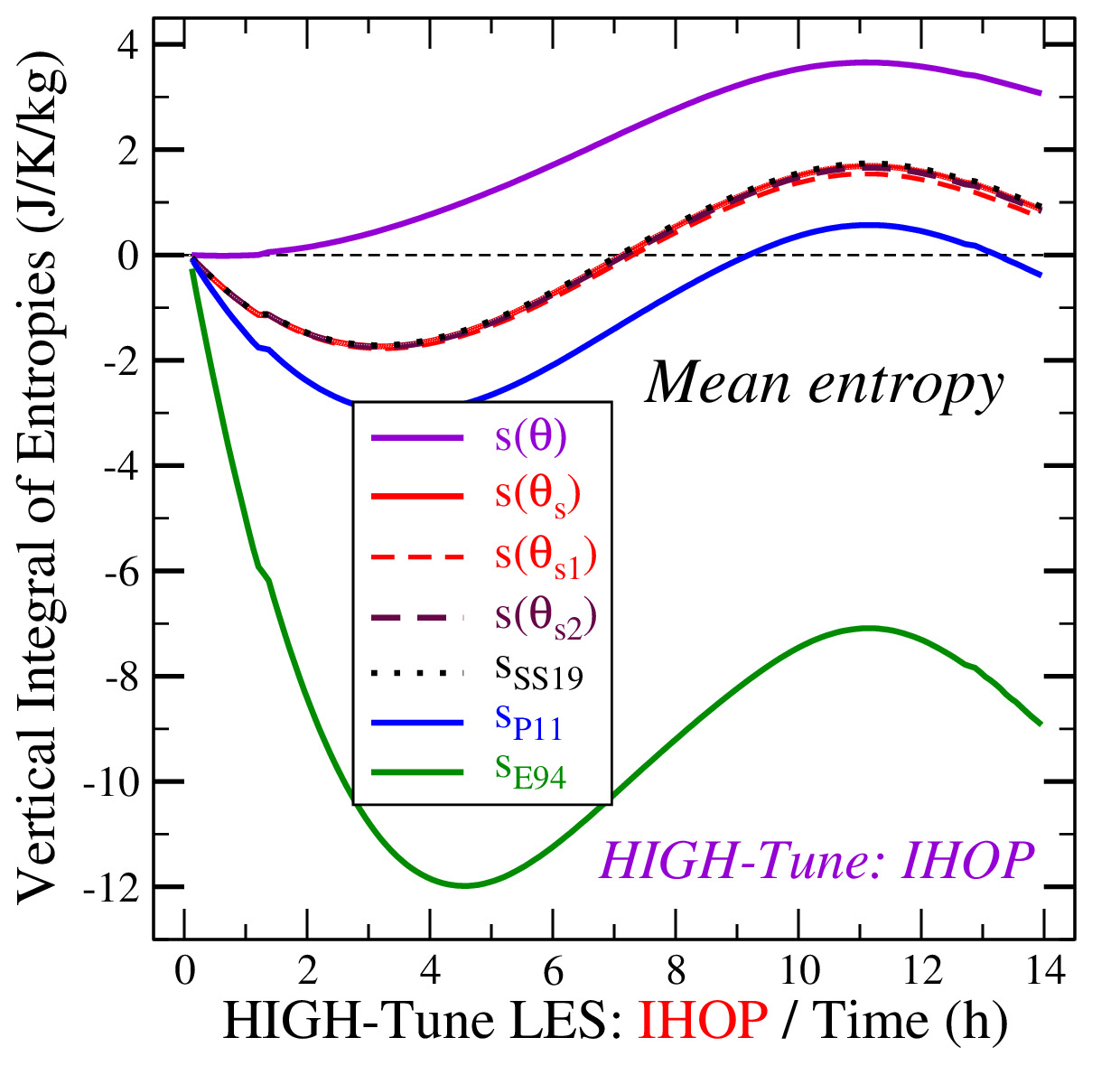} \\
\vspace*{-1.5mm}
             (\textbf{c})
\caption{Vertical mean values of entropies (J~K${}^{-1}$~kg${}^{-1}$) for three LES 
of HIGH-Tune for:
(\textbf{a}) RICO; (\textbf{b}) FIRE-I; (\textbf{c}) IHOP.
Dry air entropy with $\theta$ (solid purple); 
third law values with $\theta_s$ (solid red), 
$(\theta_s)_1$ (dashed red), 
and $(\theta_s)_2$ (long dashed dark purple);
P11 value with $\theta_e$ (solid blue);
SS19 value with $\theta_e$ (dotted black);
E94 value with $\theta_e$ (solid green).
}
\label{Figure_LES_intS}
\end{figure}

In order to specify these impacts, the entropy fluxes plotted in Figure~\ref{Figure_ARPEGE_LH_SH_all} are those calculated for ARPEGE with Equations~(\ref{eq_SH}-\ref{eq_LH}) for the fluxes of energy $SH$ and $LH$ and Equations~(\ref{eq_FSH}-\ref{eq_FLH}) for the flux of entropy $SH/T+LH/T$ (Standard) and $FS$ (New).

The new third-law formulation for the surface flux of entropy is systematically smaller than the one commonly used.
The global average decreases from $0.327$ to $0.247$~W~K${}^{-1}$~m${}^{-2}$, which represents a decrease of $0.08$~W~K${}^{-1}$~m${}^{-2}$ or $-24.5$~\% that confirms previous estimates.
Such a decrease of about 
$80$~mW~K${}^{-1}$~m${}^{-2}$
is important because it represents $16$~\% of the net surface production for entropy estimated at
$505$~mW~K${}^{-1}$~m${}^{-2}$ by \citeauthor{Bannon_2015} \cite{Bannon_2015}.
According to \citeauthor{Bannon_2015}'s estimates, this decrease of about $80$~mW~K${}^{-1}$~m${}^{-2}$ for the surface entropy production should increase the entropy production by the atmosphere from $774$ to $854$~mW~K${}^{-1}$~m${}^{-2}$, with an increase of more than $10$~\%.
This effect may not be negligible, and it may constitute a test of the properties induced by the third law.







\subsection{Entropy changes and fluxes for HIGH-Tune LES} 
\label{Results_LES_HIGHTune}

The global impacts described in the previous section based on the results of ARPEGE may depend on the more or less realistic physical parameterizations used in this NWP model (radiation, turbulence, convection, micro-physics). 

A complementary and more precise study can be carried out with the Meso-NH model outputs in LES mode and for three different cases that were studied during the HIGH-Tune project.
The RICO field study is a cumulus case over sea and off the Caribbean islands of Antigua and Barbuda, from November 2004 to Juanuary 2005 \cite{Rauber_al_2007} \cite{vanZanten_2011}.
The FIRE-I case is a stratocumulus diurnal cycle over sea and off the coast of California during July 1987 \cite{Duynkerke_al_2004}.
The IHOP case is a simulation to test the behaviour of water vapour in a clear-air (no cloud) growing convective boundary-layer (CBL) over land in Kansas, Oklahoma and Texas (USA / June 2002) \cite{Couvreux_al_2005}. 

The temporal evolutions (difference from initial values) of the vertical averages of the entropies are plotted in Figure~\ref{Figure_LES_intS}~(a,b,c) for the three LES.
As in Figure~\ref{Figure_ARPEGE_vertical_mean_s}~(a), the third-law formulations ($s_{abs}$, $s_{abs/2}$, $s_{SS19}$) coincide almost perfectly, with only a (very) small difference for the first-order approximation $s_{abs/1}$ (dashed red lines).

We also find for RICO and IHOP the same $2/3$ positioning of the curve for $s_{abs}(\theta_s)$ between those for $s_{dry}(\theta)$ and for $s_{P11}(\theta_e)$. 
This property is not as well verified in the FIRE case, where if it is verified locally in the PBL and in the free atmosphere above, the effects of the joint but opposite evolutions of these two different regions compensate each other in a complex way.
For the IHOP case in (c) we see that only $s_{dry}(\theta)$ never decreases, which shows that the various formulations are far from leading to the same results concerning the evolution of entropy for such atmospheric simulations.
As for the evolution of $s_{E94}$, it differs from all the others by much higher and faster variations.

\begin{figure}[H]
\centering
\includegraphics[width=6.6 cm]{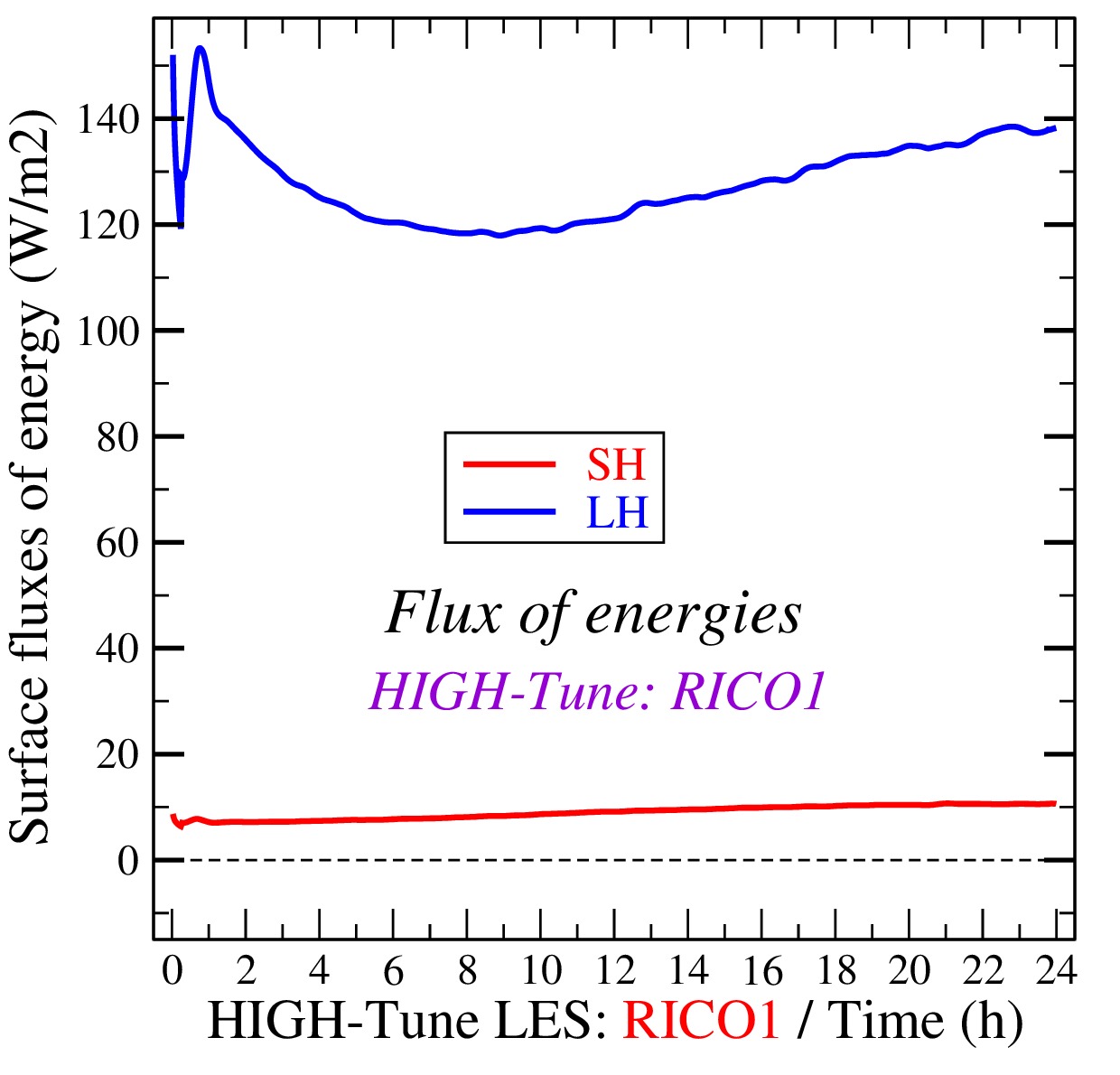}
\includegraphics[width=6.6 cm]{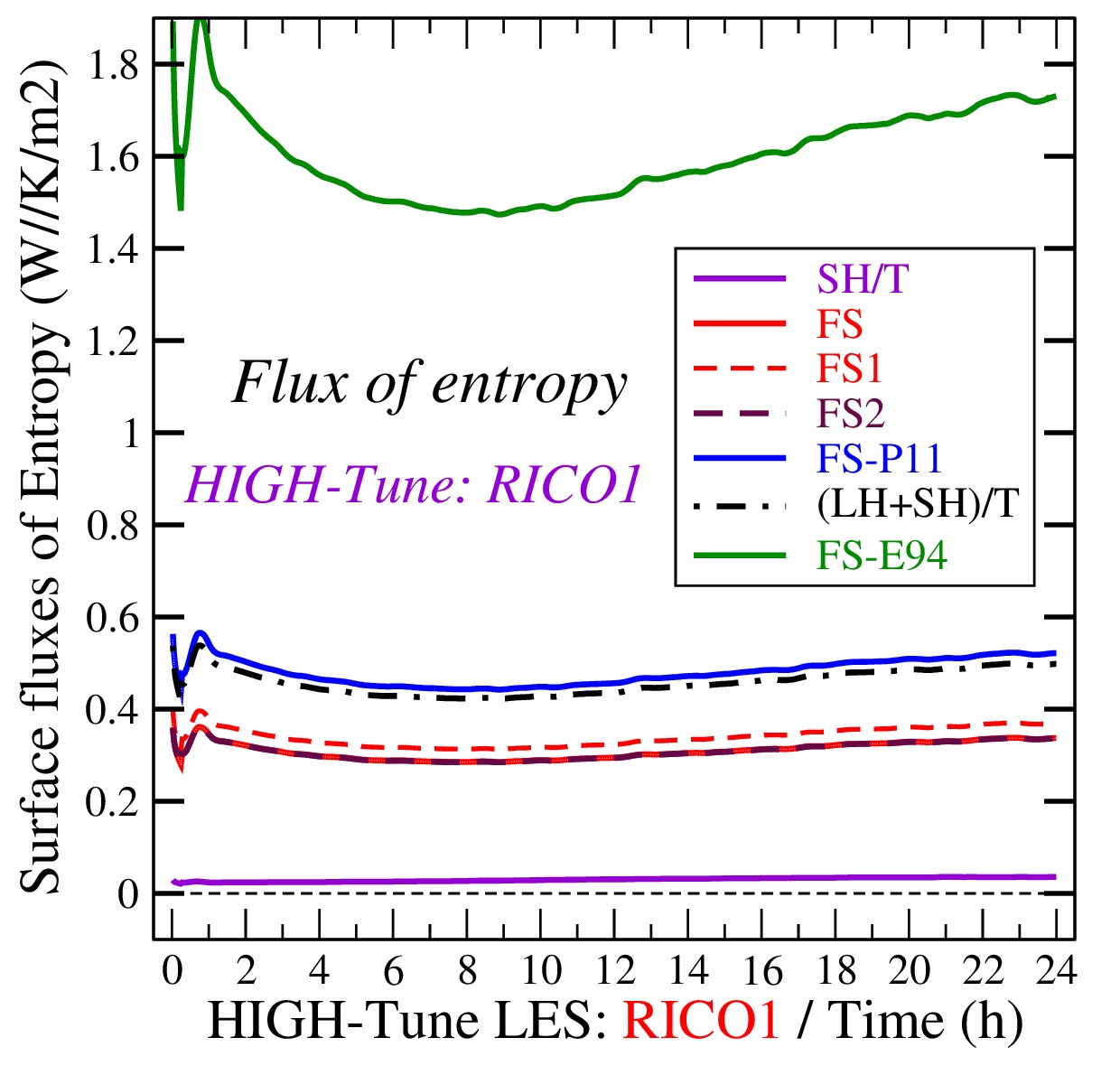} \\
\vspace*{-1.5 mm}
(\textbf{a}) \hspace*{6 cm} (\textbf{b}) \\
\includegraphics[width=6.6 cm]{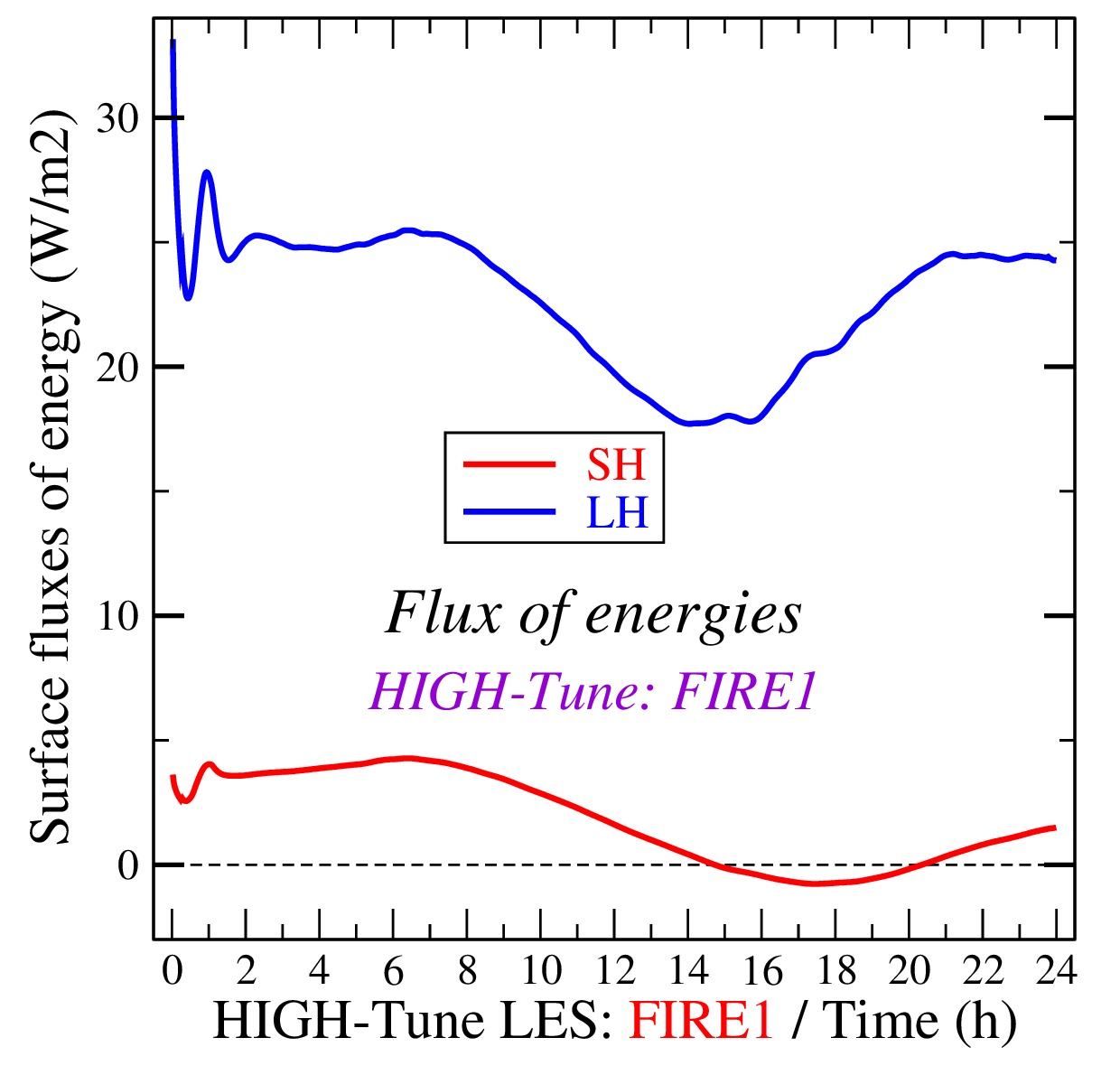}
\includegraphics[width=6.6 cm]{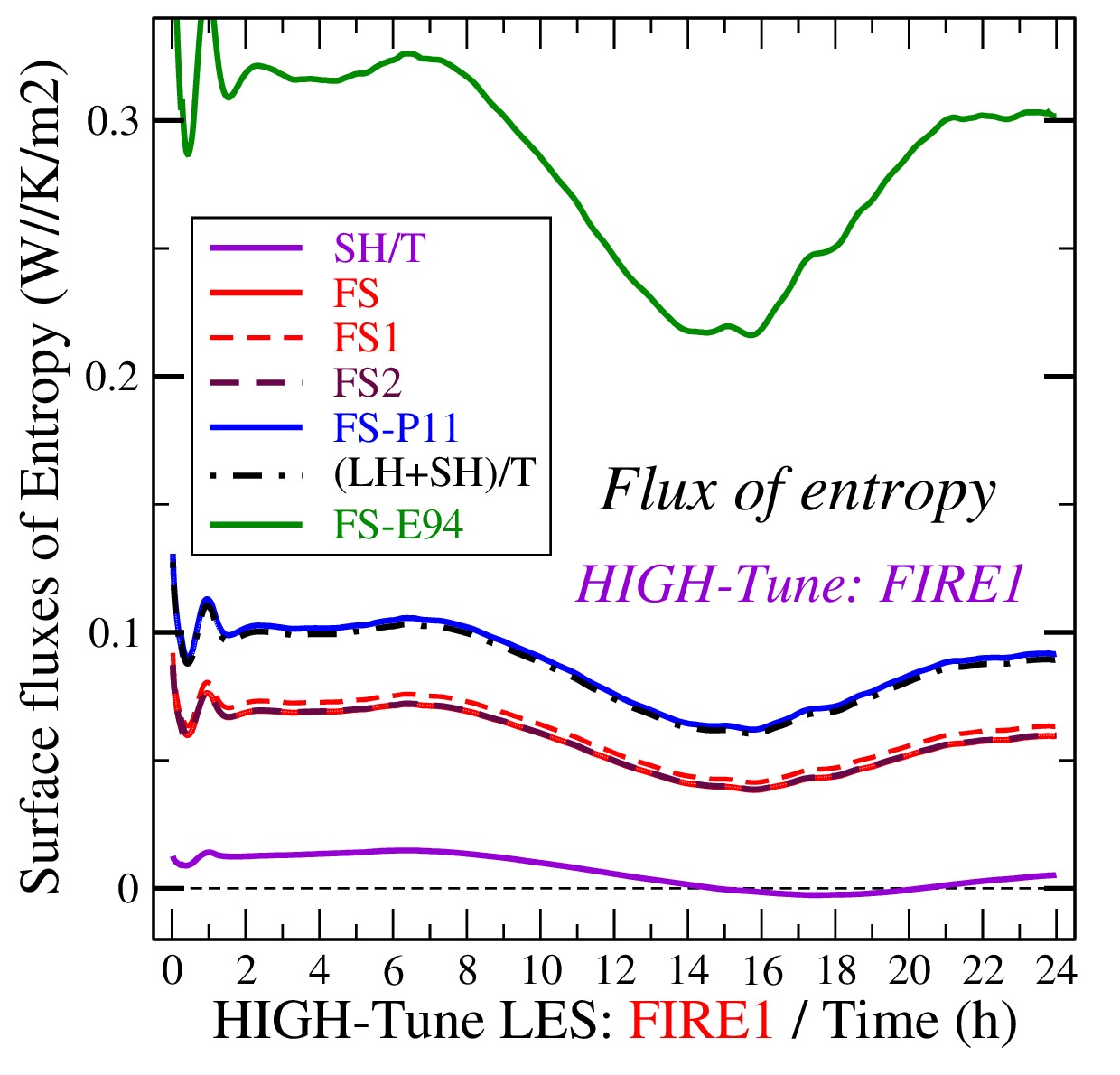} \\
\vspace*{-1.5 mm}
(\textbf{c}) \hspace*{6 cm} (\textbf{d}) \\
\includegraphics[width=6.6 cm]{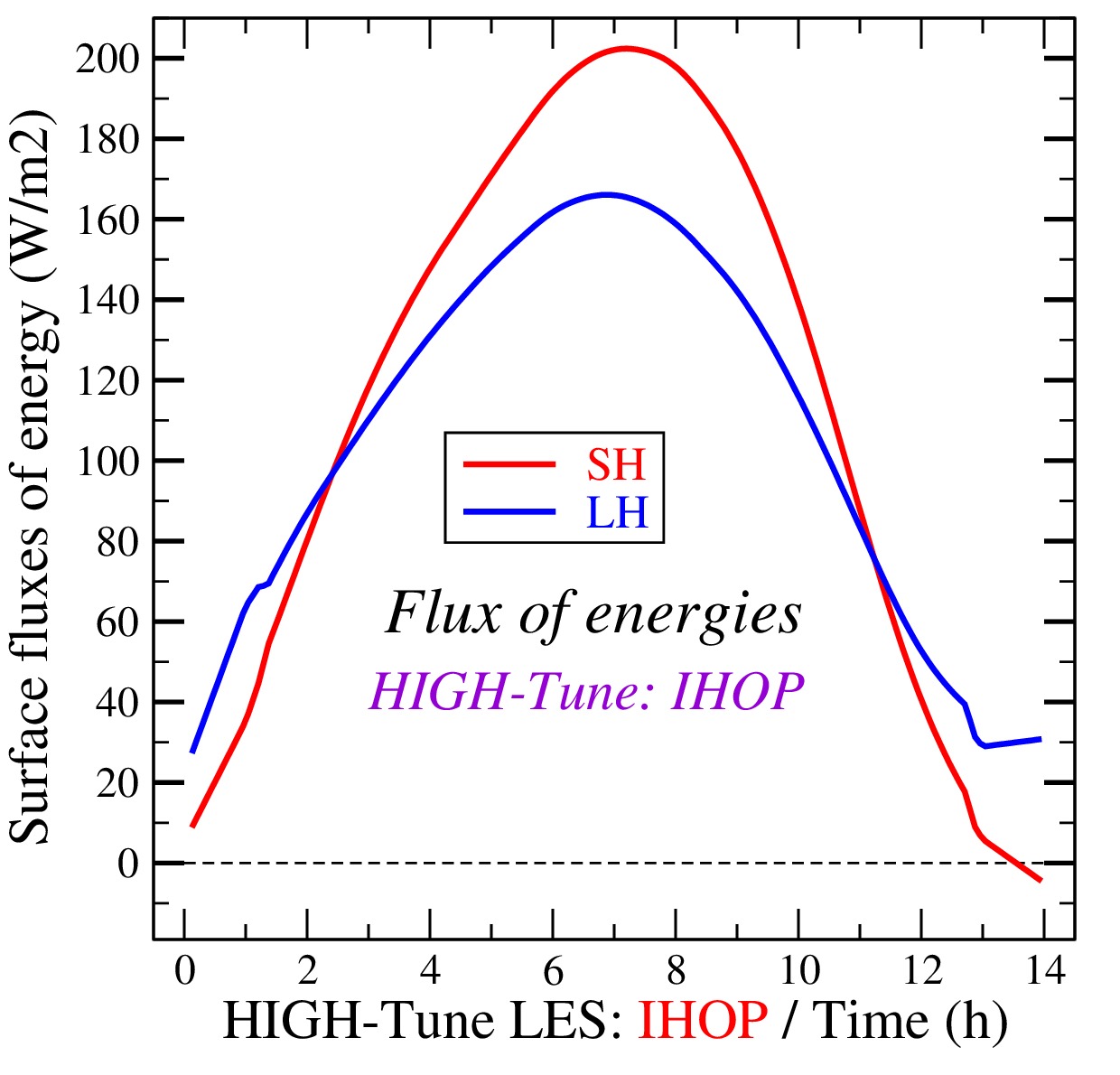}
\includegraphics[width=6.6 cm]{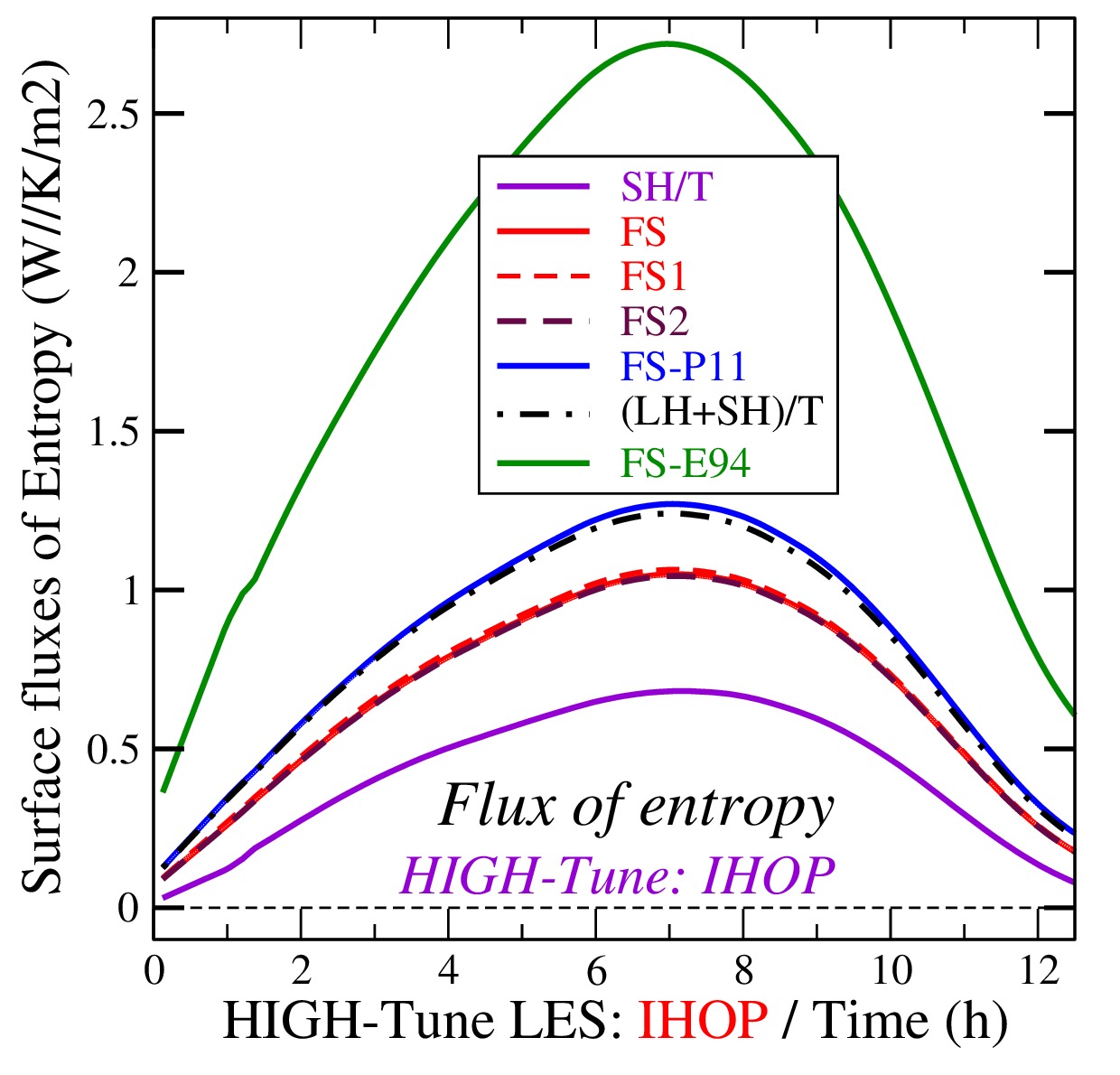} \\
\vspace*{-1.5 mm}
(\textbf{e}) \hspace*{6 cm} (\textbf{f})
\caption{Surface fluxes ($>0$ upward) computed for three LES of HIGH-Tune.
Surface fluxes of energy ($SH$ and $LH$, W~m${}^{-2}$) for:
(\textbf{a}) RICO, (\textbf{c}) FIRE-I, (\textbf{e}) IHOP.
Surface fluxes of entropy (W~K${}^{-1}$~m${}^{-2}$) for:
(\textbf{b}) RICO, (\textbf{d}) FIRE-I, (\textbf{f}) IHOP.
Dry air entropy flux $SH/T$ (solid purple); 
third law values with $\theta_s$ (solid red), 
$(\theta_s)_1$ (dashed red), 
and $(\theta_s)_2$ (long dashed dark purple);
P11 value (solid blue);
standard value $SH/T+LH/T$ (dashed dotted black);
E94 value (solid green).
}
\label{Figure_LES_LH_SH_FS}
\end{figure}

The surface fluxes of energy and entropy are plotted in Figure~\ref{Figure_LES_LH_SH_FS} for the thee LES, and the surface fluxes of entropy averaged over the simulations are shown in the Table~\ref{Table_LES_FS}.
The three cases RICO, FIRE and IHOP scan various conditions for surface fluxes.
However, common features are observed and seem to be robust and consistent with the global results obtained with ARPEGE.

\begin{table}[H]
\caption{Mean surface fluxes of entropies (W~K${}^{-1}$~m${}^{-2}$) for three of the LES of HIGH-Tune (RICO, FIRE, IHOP).
Fluxes for: $s_{dry}$ ($SH/T$); 
$s_{abs}$; $s_{abs/2}$; $s_{abs/1}$; 
$s_{P11}$; $(SH+LH)/T$; and $s_{E94}$.
}
\centering
\begin{tabular}{cccccccc}
\toprule
    \textbf{LES-cases}	&
\textbf{$SH/T$} & 
\textbf{$FS_{abs}$} & 
\textbf{$FS_{abs/2}$} & 
\textbf{$FS_{abs/1}$} & 
\textbf{$FS_{P11}$} &
\textbf{$(SH+LH)/T$} &
\textbf{$FS_{E94}$}
\\
\midrule
RICO &
$0.030$ & 
$0.310$ & 
$0.309$ & 
$0.340$ &
$0.480$ &
$0.459$ &
$1.60$
\\
FIRE & 
$0.006$ & 
$0.057$ & 
$0.057$ & 
$0.061$ &
$0.087$ &
$0.085$ &
$0.28$
\\
IHOP & 
$0.387$ & 
$0.621$ & 
$0.618$ & 
$0.632$ &
$0.762$ &
$0.744$ &
$1.68$
\\
\bottomrule
\end{tabular}
\label{Table_LES_FS}
\end{table}

Values for absolute entropy are very close for $s_{abs}$ and $s_{abs/2}$, with differences remaining small with the first order approximation $s_{abs/1}$ (dashed red lines).
This means that the convenient approximate formula 
$FS{abs} \approx FS{abs/1} \approx (SH+0.7LH)/T$ 
is verified with a good approximation for the mean circulation.

Values of $FS_{E94}$ for $s_{E94}$ are much larger than the others, and the curves and mean values for $s_{abs}$ are in a $2/3$ positioning  between those for $s_{dry}$ ($SH/T$) and for $s_{P11}(\theta_e)$ ($FS_{P11}$).

We observe another result that is a priori more surprising: $FS_{P11}$ (blue solid lines) remains very close to the commonly used value $(SH+LH)/T$ (black dashed-dotted lines).
This mystery can be solved by calculating $FS_{P11}$ from Equation~(\ref{eq_s_P11}), which implies that  
\begin{align}
 FS_{P11} \: = \:
 \overline{\rho} \:
 \overline{w' s'_{P11}}
 & \approx  \: 
 \overline{\rho} \:
  \overline{w' s'_{abs}}
 + 
 \left( s_{d}^0 - s_{l}^0 \right) \: 
  \overline{\rho} \:
  \overline{w' q'_t}
 \label{eq_FP11_old1} \\
 & \approx \: 
 \frac{T}{\theta} \:
 \frac{SH}{T}
 + 
 \left[ \: 
   \frac{c_{pd} \: T \: \Lambda}{L_v}
   + 
   \frac{T \: \left( s_{d}^0 - s_{l}^0 \right)}{L_v}
 \: \right] \:
 \frac{LH}{T} \: .
 \label{eq_FP11_old2}
\end{align}

The two bracketed terms in equation~(\ref{eq_FP11_old2}) can be evaluated with
$s_{d}^0$ and $s_{l}^0$ given by 
Equations~(\ref{eq_sd_0}) and (\ref{eq_sl_0}),
$\Lambda \approx 6$, 
$T \approx 290$~K and
$L_v \approx 2.5\:10^6$ ~J~kg${}^{-1}$,
leading to 
${c_{pd} \: T \: \Lambda}{L_v} \approx 0.7$ 
and $290 \times 
(6775-3517)/2.5 \, 10^6
\approx 0.38$, the sum of the two terms being equal to $1.08$.
These calculations indicate that $FS_{P11}$ must be a few percent larger than the commonly used value $(SH+LH)/T$.
This is true here for the three HIGH-Tune LES, both for the dashed dotted black and solid blue curves in Figures~\ref{Figure_LES_LH_SH_FS}~(b, d, f) and for the average values in Table~\ref{Table_LES_FS}. 
However, this coincidence should not be interpreted as a reinforcement of the interest of $FSP11$.
Differently, this coincidence results from a compensation of two terms, only the first of which is in agreement with the third law, since the presence of the second results from the assumption $s_d^0=s_l^0=$0 made in P11 \cite{Pauluis_2011} which contradicts the third law.

\subsection{Surface fluxes of entropy for EBEX-2000} 
\label{Results_EBEX}

As the LES are simulations that make certain assumptions and use physical parameterizations (radiation, microphysics, near-ground turbulence), it cannot be excluded that the effects described so far may be consequences of these physical assumptions and parameterizations.
It is therefore useful to conclude this study by comparing some of the results with evaluations made with observed data.

\begin{figure}[H]
\centering
\includegraphics[width=7.7 cm]{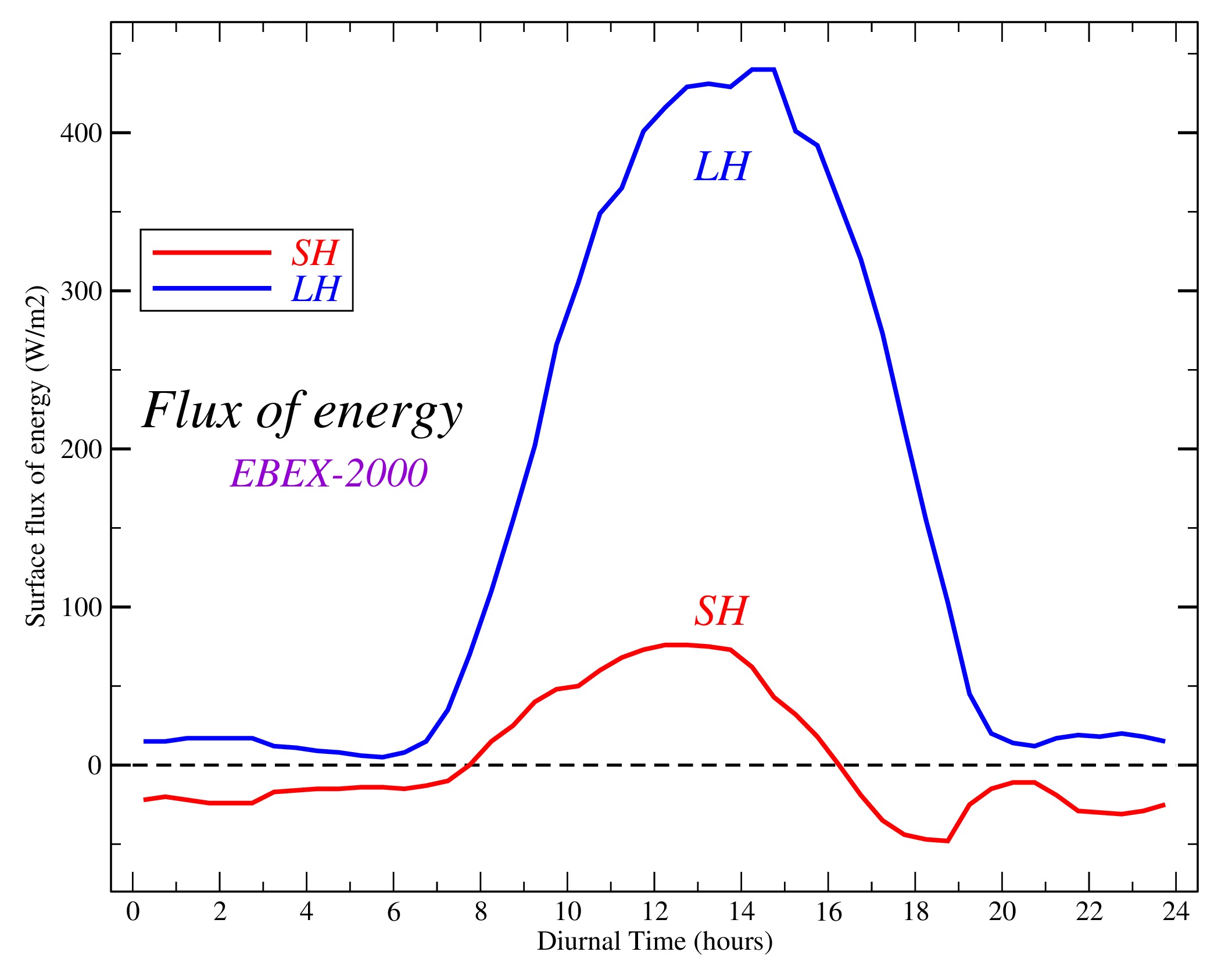}
\includegraphics[width=7.7 cm]{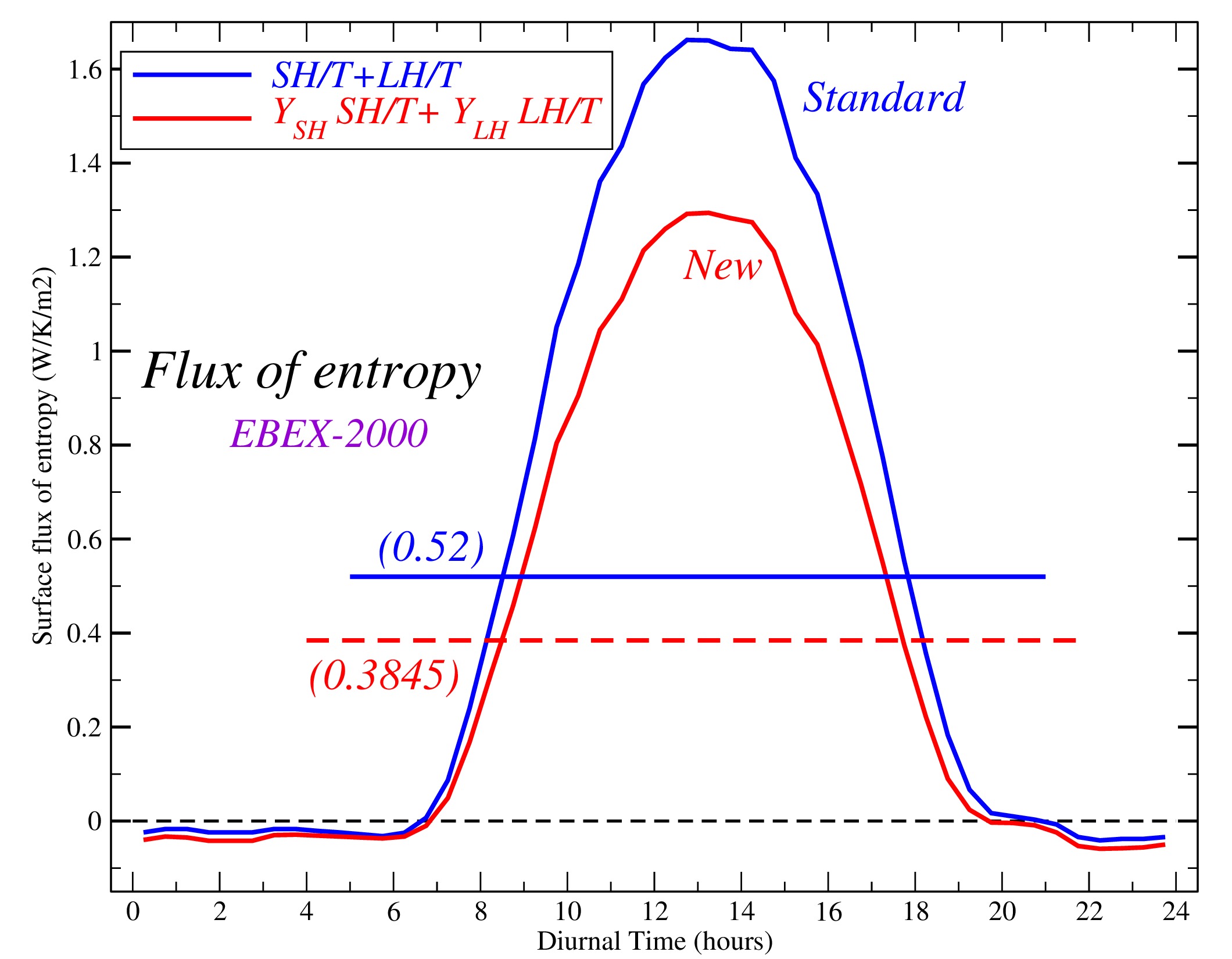} \\
\vspace*{-1.5 mm}
(\textbf{a}) \hspace*{7 cm} (\textbf{b})
\caption{Surface fluxes ($>0$ upward) computed with the 
EBEX-2000 dataset \cite{Oncley_al_2007}.
Mean diurnal cycle for about one month and 9 stations.
(\textbf{a}) 
The mean surface sensible and latent heat fluxes of energy $SH$ and $LH$ 
(W~m${}^{-2}$).
(\textbf{b}) 
The mean standard surface fluxes of entropy 
$SH/T+LH/T$
compared to the new version 
$FS = Y_{SH} \: SH/T + Y_{LH} \: LH/T
 \approx SH/T + 0.7 \: LH/T $ 
   (W~K${}^{-1}$~m${}^{-2}$). 
}
\label{Figure_EBEX_LH_SH}
\end{figure}   

Data for the EBEX-2000 campaign \cite{Oncley_al_2007} are averaged over the period between July 28 and August 26, 2000 for 9 observation sites over a flood-irrigated cotton field of $1.600$~m times $800$~m large  in the San Joaquin Valley of California, USA.

The surface entropy fluxes shown in Figure~\ref{Figure_EBEX_LH_SH}~(b) confirm the impact of the third law of thermodynamics, with the formula for $FS$ given by Equation~(\ref{eq_FS}) and with the factor close to $0.7$ in front of $LH/T$. 
The entropy flux goes from $0.52$~W~K${}^{-1}$~m${}^{-2}$ for $SH/T+LH/T$ to $0.3845$~W~K${}^{-1}$~m${}^{-2}$ for $FS$. 
This corresponds to a decrease of $-0.136$~W~K${}^{-1}$~m${}^{-2}$ or $-26$~\%.
These impacts are of the same order of magnitude as those indicated at a larger scale of space and time by the ARPEGE and Meso-NH models.

\section{Discussion and conclusion} 
\label{Conclusion}

One of the consequences of the absolute definition of atmospheric entropy is to lead to a formulation of turbulent flows at the surface that seems different from the one used until now.
The usual formulation $SH/T+LH/T$ seems to be replaced by a formula close to $SH/T + 0.7\: LH/T$, with a decrease in this turbulent entropy flux of about $-80$~mW~K${}^{-1}$~m${}^{-2}$ or $-25$~\% in global average.
This decrease in surface entropy flux could lead to an increase in the necessary atmospheric entropy production of about $+10$\:\%, which may not be negligible in studies that evaluate this entropy production for all irreversible processes in the atmosphere. 

More generally, the desire and possibility of calculating the entropy of the atmosphere through its absolute formulation must have other consequences that could be interesting to address later.
This is the case for the impact of the so-called "diabatic" terms, which are created by the absorption and emission of radiation as well as by changes in the phases of water.
Indeed, Equation~(\ref{eq_s_turb}) can be written, neglecting some terms due to dissipation for example, such as
\begin{align} 
 \overline{\rho} \:
       \left[
 \frac{\partial \, \overline{s}}{\partial t}
 + \,
 \left( \overline{\vec{u}} . \vec{\nabla} \right)
 \overline{s}
        \right]
 + \,
 \vec{\nabla} . 
 \left( \overline{\rho} \, \overline{ s \, \vec{u}} \right)
 \: = \:
         \frac{\overline{\rho \, \dot{Q_R}}}{T}
 \, - \, \overline{
          \frac{\rho}{T}
          \sum_{k=0}^3  \: \mu_k \: 
          \frac{d_e q_k}{dt}
         } 
 \, - \, \overline{
          \frac{\rho}{T}
          \sum_{k=1}^3 \: \mu_k \: 
          \frac{d_i q_k}{dt}
         } 
 \label{eq_s_turb_2} 
\end{align} 
where $- \, \overline{\vec{\nabla} . \vec{J}_R}$ is symbolically replaced by the entropy change 
$\overline{\rho \, \dot{Q_R}/T}$ due to radiation and
$ - \, \overline{\vec{\nabla} . \vec{J}_s}$ 
by the usual Gibbs summations for the product of chemical potentials ($\mu_k$) and changes in the specific contents 
($dq_k/dt = d_eq_k/dt + d_iq_k/dt$)
 over $k=0,1,2,3$ for dry air, water vapour, liquid water and ice.

This study emphasizes that the sensible and latent heat turbulent fluxes must be calculated with the opposite of the term
$ \vec{\nabla} . \left( \overline{\rho} \, 
 \overline{ s \, \vec{u}} \right)$
in the left-hand side of Equation~(\ref{eq_s_turb_2}), and not by any of the two sums in the right-hand side.

The other important point is that if the phase changes are reversible, i.e. if they occur with the same chemical potentials ($\mu_k$), then the last sum over water species in Equation~(\ref{eq_s_turb_2}) is exactly zero and there is no entropy production.
Thus, the impact of changes of phase on entropy production in the atmosphere should only be calculated for irreversible processes such as supercooled water, mixed liquid/solid phases, or the creation of liquid or solid precipitations that falls toward the surface.
This vision is very different from the one with dry air entropy and $ds(\theta)/dt$, where latent heat release due to phase changes, even reversible, are sources of $s(\theta)$, while they do not create any entropy. 

It is of course always possible to transform Equation~(\ref{eq_s_turb_2}) for entropy into equations for the temperature $T$ or for the potential temperature $\theta$, which are the equations used in all forecast models.
This rewriting means transferring to the right-hand side all terms that do not explicitly depend on changes in $T$ or $\theta$.
But the constraint imposed on the atmosphere by the imbalance in entropy  fluxes of radiation at its upper and lower boundaries is on the changes in moist-air entropy $ds/dt$ or $d\theta_s/dt$, not on the dry air version $d\theta/dt$, nor $dT/dt$.
Therefore, the possible principle of maximum entropy production (MaxEP), or even the maximization of entropy itself, depend on how the entropy is defined, as shown in this study with the visibly very different formulations for the entropy flux at the surface and for the vertical integral of the entropy. 

In the same vein, this study confirms that the vision of P11 and E94 which depends on the equivalent potential temperature $\theta_e$ would overestimate the temporal variations as well as the vertical and meridian flows of the absolute entropy and $\theta_s$.
This overestimation of the impact of changes of $q_t$ by a factor of about $1/3$ may have consequences on the calculations of entropy production, if entropy is defined with assumptions different from those recommended by the third law of thermodynamics.

The controversy over the possibility of defining the entropy of the atmosphere by following or disregarding the third law recommendations seems to be understood in a new way by this study.
Indeed, we can consider with Max Planck that the same rules must apply to radiation as to matter in order to be able to study a coherent open system consisting of the surface and atmosphere that are illuminated by the sun from space. 
The consequence is the use of Boltzmann's formula $S = k_B \: \ln(W)$ with no additional term for both radiation and matter, with a radiation entropy equal to $(4/3) \: \sigma_B \: T^3$ and a material entropy whose reference values must be in accordance with  the third law.


\abbreviations{The following abbreviations are used in this manuscript:\\
\noindent 
\begin{tabular}{@{}ll}
ARPEGE & Action de Recherche Petite \'Echelle Grande 
    \'Echelle\\
EBEX & Energy Balance Experiment\\
EOL & Earth Observing Laboratory\\
FIRE & First ISCCP Regional Experiment\\ 
GCM & General Circulation Model\\
IAPWS & International Association for the Properties 
       of Water and Steam\\
IHOP & International H${}_{2}$O Project\\
ISCCP & International Satellite Cloud Climatology Project\\
JANAF & Joint Army Navy NASA Air Force\\
LES & Large Eddy Simulation\\
LH & Latent Heat (flux)\\
LW & Long Wave (radiation)\\
MPI & Max Planck Institute\\
NCAR & National Center for Atmospheric Research \\
NIST & National Institute of Standards and Technology\\
NSF & National Science Foundation\\
NWP & Numerical Weather Prediction\\
PBL & Planetary Boundary Layer\\
RICO & Rain in Cumulus over the Ocean\\
SH & Sensible Heat (flux)\\
SIA &  Seawater Ice Air\\
SW & Sort Wave (radiation)\\
TEOS & Thermodynamic Equation of Seawater\\
TOA & Top Of Atmosphere \\
\end{tabular}
}

\section{funding}
\funding{The LES simulations were made available by Fleur Couvreux and this work received funding from Grant HIGH-TUNEANR-16-CE01-0010.
The EBEX-2000 data are provided by NCAR/EOL under the sponsorship of the NSF. \url{https://data.eol.ucar.edu/}}



\section{conflicts of interest}
\conflictsofinterest{The authors declare no conflict of interest.}

\appendixtitles{yes}
\appendix
\section{The third-law moist-air entropy}

The moist-air entropy is computed with the Gibbs's proposition as a weighted sum of partial entropies for dry air ($s_d$), water vapour ($s_v$), liquid water ($s_l$) and ice ($s_i$) (see \cite[hereafter HH87]{Hauf_Holler_1987}, 
\cite[hereafter E94]{Emanuel94}, \cite{Pauluis_al_2010}, \cite[hereafter P11]{Pauluis_2011}, \cite{Zdun_Bott_2004}, \cite[hereafter M11]{Marquet_2011} and \cite[hereafter SS19]{Stevens_Siebesma_2019}), leading to
\begin{align}
 s  & \: = \: 
 q_d \: s_d \: + \: q_v \: s_v \: + \:
 q_l \: s_l \: + \: q_i \: s_i 
  \label{eq_s_abs_sum}
  \: .
\end{align}
The specific contents ($q_d$, $q_v$, $q_l$, $q_i$) are the mass of species per unit mass of moist air (i.e. the concentrations).
The entropy of each species are computed in terms of the temperature ($T$), the total pressure ($p$) and the water-vapour partial pressure ($e$), leading to
\begin{align}
 s_d(T,p-e)  & \: = \:
  c_{pd} \: \ln\left(\frac{T}{T_0}\right) 
  \: - \: R_{d} \: \ln\left(\frac{p-e}{p_0}\right)
  \: + \: s_d^0
  \label{eq_sd} \: , \\
 s_v(T,e)  & \: = \:
  c_{pv} \: \ln\left(\frac{T}{T_0}\right) 
  \: - \: R_{v} \: \ln\left(\frac{e}{p_0}\right)
  \: + \: s_v^0
  \label{eq_sv} \: , \\
 s_l(T)  & \: = \:
  c_l \: \ln\left(\frac{T}{T_0}\right) 
  \: + \: s_l^0
  \label{eq_sl} \: , \\
 s_i(T)  & \: = \:
  c_i \: \ln\left(\frac{T}{T_0}\right) 
  \: + \: s_i^0
  \label{eq_si} \: ,
\end{align}
where $T_0=273.15$~K and $p_0=1000$~hPa are standard constant values.
The gas constants 
($R_d \approx 287.06$~J~K${}^{-1}$~kg${}^{-1}$, 
$R_v \approx 461.53$~J~K${}^{-1}$~kg${}^{-1}$) 
and the specific heats at constant pressure
($c_{pd}\approx 1004.7$~J~K${}^{-1}$~kg${}^{-1}$, 
$c_{pv}\approx 1846.1$~J~K${}^{-1}$~kg${}^{-1}$, 
$c_l\approx 4218$~J~K${}^{-1}$~kg${}^{-1}$, 
$c_i\approx 2106$~J~K${}^{-1}$~kg${}^{-1}$) 
are assumed to be constant in the atmospheric range of temperatures (say between $200$~K and $330$~K).

The integration constant $s_d^0$, $s_v^0$, $s_l^0$ and $s_i^0$ in Equations~(\ref{eq_sd}) to (\ref{eq_si}) are four unknown reference entropies.
They are not all independent, because the entropies of water species are linked by the relations
\begin{align}
 s_l  & \: = \: s_v
  - \frac{L_v(T)}{T} \: + R_v \: \ln(H_l)
  \label{eq_sl_sv} \: , \\
 s_i  & \: = \: s_v
  - \frac{L_s(T)}{T} \: + R_v \: \ln(H_i)
  \label{eq_si_sv} \: ,
\end{align}
where $L_v(T)$ and $L_s(T)$ are the latent heat of vaporisation and sublimation.
The relative humidities $H_l=e/e_{sw}$ and $H_i=e/e_{si}$ depend on the saturating pressure with respect to liquid water $e_{sw}(T)$ and ice $e_{si}(T)$, respectively.
Therefore, only the dry-air value $s_d^0$ and one of the water values must be known if $L_v(T)$, $L_s(T)$, $H_l(T,e)$ and $H_i(T,e)$ are known.
 
According to HH87 \citep[p. 2891]{Hauf_Holler_1987}, the (absolute) "{\em values of the zero entropies\/}" (i.e. $s_d^0$, $s_v^0$, $s_l^0$ and $s_i^0$) "{\em have to be determined experimentally or by quantum statistical considerations\/}".
These methods refer to the so-called third law of thermodynamics, which consists in adopting $s=0$ for the most stable form of any solid at absolute temperature $T=0$~K
(\cite{Nernst_1906},
\cite{Nernst_1907},
\cite{Planck_1917},
\cite{Lewis_Randall_1923},
\cite{Simon_1927},
\cite{Kelley_1932},
\cite{Giauque_1949},
\cite{Tiselius_1949},
\cite{Simon_1951},
\cite{Schrodinger_1952},
\cite{Lewis_Randall_1961},
\cite{Gokcen_Reddy_1996},
\cite{Bejan_2016},
\cite{Masanes_Oppenheim_2017}).

The experimental method is based on measurements of $c_p(T)$ and an integration of $c_p(T)/T$ from zero Kelvin to the standard temperature $T_0=273.15$~K for the solid, liquid and vapour states of all chemical elements.
It is also necessary to add the values $L_k/T_k$ for all changes of phase occurring at the transition temperatures $T_k$ with a latent heat $L_k$.
The method based on quantum statistics gives the same results and is an application of the Boltzmann-Planck equation $s = k \: \ln(W)$, with the need to calculate $W$ for the translation, rotation and vibration degrees of freedom of atoms and molecules (\cite{Boltzmann_1877},
\cite{Planck_1917},
\cite{Lewis_Randall_1923},
\cite{Gordon_Barnes_1932},
\cite{Kelley_1932},
\cite{Gordon_34},
\cite{Gordon_35},
\cite{Lewis_Randall_1961},
\cite{Gokcen_Reddy_1996}).

The absolute entropies $s_d^0$, $s_v^0$, $s_l^0$ and $s_i^0$ listed in HH87 for dry air, water vapour, liquid water and ice are computed at $T_0 = 273.15$~K and $p_0=1000$~hPa, leading to
\begin{align}
 s_d^0  & \: \approx 
   \: 6775 \; \; 
          \mbox{J~K${}^{-1}$~kg${}^{-1}$}
   \; \; &&(s^{\ast}_d \: = \:
   \: 6863 \; \; 
          \mbox{J~K${}^{-1}$~kg${}^{-1}$}
   )
  \label{eq_sd_0} \: , \\
 s_v^0  & \: \approx 
   \: 10320 \; \; 
          \mbox{J~K${}^{-1}$~kg${}^{-1}$}
   \; \; &&(s^{\ast}_v \: = \:
   \: 10482 \; \; 
          \mbox{J~K${}^{-1}$~kg${}^{-1}$}
   )
  \label{eq_sv_0} \: , \\
 s_l^0  & \: \approx 
   \: 3517 \; \; 
          \mbox{J~K${}^{-1}$~kg${}^{-1}$}
   \; \; &&(s^{\ast}_l \: = \:
   \: 3886.4 \; \; 
          \mbox{J~K${}^{-1}$~kg${}^{-1}$}
   )
  \label{eq_sl_0} \: , \\
 s_i^0  & \: \approx 
   \: 2296 \; \; 
          \mbox{J~K${}^{-1}$~kg${}^{-1}$}
   \; \; &&(s^{\ast}_i \: = \:
   \: 2480 \; \; 
          \mbox{J~K${}^{-1}$~kg${}^{-1}$}
   )
  \label{eq_si_0} \: .
\end{align}
These absolute entropies corresponds to those published at the MPI in SS19 \cite{Stevens_Siebesma_2019} if computed at $T_0 = 273.15$~K and $p_0=1000$~hPa ($s_d^0 \approx 6783$ and $s_v^0 \approx 10321$~J~K${}^{-1}$~kg${}^{-1}$).
The same is true for the dry air value published in \cite{Lemmon_al_2000} ($s_d^0 \approx 6772$~J~K${}^{-1}$~kg${}^{-1}$) and for the seawater in \cite{Millero_1983} ($s_l^0 \approx 3516$~J~K${}^{-1}$~kg${}^{-1}$), even if these recommendations were not subsequently retained in the IAPWS and TEOS-10 formulations (\citep{Feistel_al_2010} \citep{Feistel_2018}).

The star values are computed for the temperature of $298.15$~K and for $p_0=1000$~hPa by adding $c_{px}\:\ln(298.15/273.15)$, with $c_{px}=c_{pd}$, $c_{pv}$, $c_l$ or $c_i$ depending on the dry-air or water species.
These values are in good agreement with the thermochemical values (see for instance the Table~\ref{Table_S0_C98} for $s^{\ast}_v$ and $s^{\ast}_l$).
The difference between the dry air values is discussed in \citep{Marquet_2017}.
The larger value of $s^{\ast}_d$ ($+2.3$~\%) considered in HH87 \citep{Hauf_Holler_1987} might be explained by taking into account the solid phase change $\alpha$-$\beta$ occurring at $23.85$~K for O${}_2$, forming a kind of Dirac pulse for $c_p(T,p_0)$, but without latent heat \citep{Fagerstroem_1969}. 


\begin{table}[H]
\caption{Absolute entropies $s^{\ast}$ at $298.15$~K and $1000$~hPa for the main atmospheric species given by the NIST-JANAF (fourth edition) thermochemical tables \cite{Chase_98} (J~K${}^{-1}$~kg${}^{-1}$).
}
\centering
\begin{tabular}{cccccccc}
\toprule
    \textbf{Absolute entropies}	&
\textbf{N${}_2$} & 
\textbf{O${}_2$} & 
\textbf{Ar} & 
\textbf{CO${}_2$} & 
\textbf{(Dry-air)} &
\textbf{H${}_2$O(vap)} &
\textbf{H${}_2$O(liq)}
\\
\midrule
$s^{\ast}$ (J~K${}^{-1}$~kg${}^{-1}$) &
$6840$ & 
$6411$ & 
$3876.16$ & 
$4858$ &
$(6701)$ &
$10482$ &
$3883$
\\
    accuracy & 
$\pm 1.4$ & 
$\pm 2.2$ & 
$\pm 0.08$ & 
$\pm 2.7$ &
$(\pm 1.6)$ &
$\pm 2.3$ &
$\pm 4.4$
\\
\bottomrule
\end{tabular}
\label{Table_S0_C98}
\end{table}

Few papers have studied the specific entropy of moist air $s_{abs}$ using absolute values based on the third law.
This is the case for the three papers HH87 \cite{Hauf_Holler_1987}, M11 \citep{Marquet_2011} and SS19 \cite{Stevens_Siebesma_2019}, which lead to the same formulation
\begin{align}
 s_{abs}  & \: = \: s_{HH87}  \: = \: s_{M11} \: = \: s_{SS19}
  \label{eq_s_abs}
  \: .
\end{align}
However, the moist-air entropy is a quantity which is not often used in meteorology and there is a tradition in atmospheric sciences of using potential temperature variables instead.
It is shown in M11 \cite{Marquet_2011} that it is possible to define and compute a potential temperature $\theta_s$ which  becomes truly synonymous with the moist-air entropy, with the use of the reciprocal properties
\begin{align}
 s_{abs}  & \: = \: 
    c_{pd} \: 
    \ln\left(\frac{\theta_s}{T_0}\right) 
    \: + \:  s_{d}^0
  \label{eq_s_thetas}
  \: , \\
  \theta_s & \: = \: T_0 \; 
   \exp\left( \: 
         \frac{s_{abs} \: - \: s_{d}^0 }{ c_{pd} } \: 
       \right)
\label{eq_thetas_s}
\: ,
\end{align}
where $T_0 = 273.15$~K, $s_{d}^0 \approx 6775$~J~K${}^{-1}$~kg${}^{-1}$ and $c_{pd} \approx 1004.7$~J~K${}^{-1}$~kg${}^{-1}$ are three constant.
Similar relationships have been suggested in HH87 \cite{Hauf_Holler_1987} to define another version of $\theta_s$ but with $s_{d}^0$ and $c_{pd}$ replaced by variable terms depending on the water content $q_t = q_v + q_l + q_i$, which prevent this other version from being synonymous with $s$ under all conditions if $q_t$ is not a constant.

The formulation of $\theta_s$ can be written as
\begin{align}
  {\theta}_{s} 
   & = \: 
    \theta \; \:
    \exp\! \left( - \: \frac{L_v \: q_l + L_s \: q_i}{c_{pd} \: T} \right)
       \:
    \exp\! \left(  \Lambda \: q_t  \right)
\nonumber \\
   &  \; \; \; \; \; \; \times \:
        \left( \frac{T}{T_0}\right)^{\!\!\lambda \,q_t}
 \!  \! \left( \frac{p}{p_0}\right)^{\!\!-\kappa \,\delta \,q_t}
 \!  \! \left( \frac{r_0}{r_v} \right)^{\!\!\gamma\,q_t}
      \frac{(1\!+\!\eta\,r_v)^{\,\kappa \, (1+\,\delta \,q_t)}}
           {(1\!+\!\eta\,r_0)^{\,\kappa \,\delta \,q_t}}
     \; {(H_l)}^{\, \gamma \, q_l} \;
     \; {(H_i)}^{\, \gamma \, q_i}
\label{eq_thetas} \: ,
\end{align}
where the so-called dry-air potential temperature is 
\begin{align}
\theta = T \: {\left( \frac{p_0}{p} \right)}^{\kappa}
\label{eq_theta} \: ,
\end{align}
with $\kappa = R_d/c_{pd} \approx 0.2857$.
It is demonstrated in M11 \cite{Marquet_2011} that this formulation of $\theta_s$ verifies the expected constraints, such as converging to the $\theta$ value for dry air ($q_t=r_v=q_l=q_i=0$).
In addition, the entropy $s_{abs}$ converges towards the expected Bauer  value \cite{Bauer_1908_inbook} valid for dry air:
\begin{align}
s_{dry} & = \: c_{pd} \: \ln(\theta) + 
 \left[\, 
     s_{d}^0 - c_{pd} \: \ln(T_0)  
 \, \right]
\label{eq_s_theta_dry} \: .
\end{align}

The other constants in Equation~(\ref{eq_thetas}) are $\eta = R_v/R_d \approx 1.608$,
$\gamma = R_v/c_{pd} \approx 0.46$,
$\delta = \eta - 1  \approx 0.608$,
$\lambda = c_{pv}/c_{pd} - 1 \approx 0.8375$.
The reference values are defined for
$e_0 = e_{sw}(T_0) \approx 6.11$~hPa and
$\eta \: r_0 = e_0/(p_0-e_0)$, leading to
$r_0 \approx 3.82$~g~kg${}^{-1}$ and 
$\Lambda =  (s_{v0} - s_{d0}) /c_{pd}  \approx 5.87$,
where 
$s_{v0} = s_v(T_0,e_0) \approx 12673$~J~K${}^{-1}$~kg${}^{-1}$
and 
$s_{d0} = s_d(T_0,p_0-e_0) \approx 6777$~J~K${}^{-1}$~kg${}^{-1}$.
These values $s_{v0}$ and $s_{d0}$ are larger than $s_v^0$ and $s_d^0$  given by in Equations~(\ref{eq_sv_0}) and (\ref{eq_sd_0}), due to the impact of the change of pressure $- R_v\, \ln(e_0/p_0)$ and $- R_d\, \ln[(p_0-e_0)/p_0]$. 
The NIST-JANAF value $s^0_d = 6701$~J~K${}^{-1}$~kg${}^{-1}$ would lead to $\Lambda \approx 6.03$, which is only $2.7$~\% larger than $5.87$.
The impact of uncertainty on the term $s^0_d$ is small for the rest of the calculations.




One of the main interests of having expressed the entropy of moist air in this form dependent on the variable ${\theta}_s$ given by Equation~(\ref{eq_thetas}) is the possibility of rigorously calculating first and second order approximations \cite{Marquet_2015_WGNE_thetas2}, leading to
\begin{align}
  {({\theta}_{s})}_1
   & = \: \theta \; \:
    \exp\! \left( - \: \frac{L_v \: q_l + L_s \: q_i}{c_{pd} \: T} \right)
       \:
    \exp\! \left(  \Lambda \: q_t  \right)
   \: = \: \theta \; \:
        \exp\! \left( 
         - \: \frac{L_v \: q_l + L_s \: q_i}{c_{pd} \: T}
         + \: \Lambda \: q_t  
          \right)
\label{eq_thetas1} \: , \\
  {({\theta}_{s})}_2
   & = \: {({\theta}_{s})}_1 \;
  {\left( \frac{r_v}{r_{\ast}} \right)}^{\! - \: \gamma \: q_t}
  \; 
  \exp\! \left[ \: - \: \gamma \: (q_l+q_i) \right]
\label{eq_thetas2} \: , \\
  {({\theta}_{s})}_2 
   & = \: \theta \;
    \exp\! \left( 
    - \: \frac{L_v \: q_l + L_s \: q_i}{c_{pd} \: T} 
    + \: \left[ \: \Lambda - \: \gamma \; 
          \ln{\left( \frac{r_v}{r_{\ast}} \right)}^{\!}
         \: \right] \: q_t  \; 
    \: - \: \gamma \: \; (q_l+q_i) 
    \right)
\label{eq_thetas2_bis} \: ,
\end{align} 
where $r_{\ast} = 0.0124$~kg~kg${}^{-1}$ is a tuning parameter (\url{https://arxiv.org/abs/1901.08108}).

\reftitle{References}
\externalbibliography{yes}
\bibliography{Marquet_MDPI_Entropy_R1.bib}
\end{document}